\documentclass[aps,superscriptaddress,nofootinbib]{revtex4}
\usepackage{hyperref}
\usepackage{epsfig,rotating}
\usepackage{amsmath,amssymb}
\usepackage{dsfont}
\usepackage{graphicx}

\numberwithin{equation}{section}

\newcommand{\p}{\omega_p}

\newcommand{\be}{\begin{equation}}
\newcommand{\ee}{\end{equation}}
\newcommand{\bea}{\begin{eqnarray}}
\newcommand{\eea}{\end{eqnarray}}

\begin{document}

\title{ Heisenberg-Langevin vs. quantum master equation }

\author{Daniel Boyanovsky}
\email{boyan@pitt.edu} \affiliation{Department of Physics and
Astronomy, University of Pittsburgh, Pittsburgh, PA 15260}
\author{David Jasnow}
\email{jasnow@pitt.edu}
\affiliation{Department of Physics and
Astronomy, University of Pittsburgh, Pittsburgh, PA 15260}
\date{\today}

\begin{abstract}
The quantum master equation is an important tool in the study of quantum open systems. It is often derived under a set of approximations, chief among them the Born (factorization) and Markov (neglect of memory effects) approximations. In this article we study the paradigmatic model of quantum Brownian motion of an harmonic oscillator coupled to a bath of oscillators with a Drude-Ohmic spectral density. We obtain analytically   the \emph{exact} solution of the Heisenberg-Langevin equations, with which we study correlation functions in the asymptotic stationary state. We compare the \emph{exact} correlation functions to those obtained in the asymptotic long time limit with the quantum master equation in the Born approximation   \emph{with and without} the Markov approximation. In the latter case we implement a systematic derivative expansion that yields the \emph{exact} asymptotic limit under the factorization approximation \emph{only}. We find discrepancies that could be significant when the bandwidth of the bath $\Lambda$ is much larger than the typical scales of the system. We study the \emph{exact} interaction energy as a \emph{proxy} for the correlations   missed by the Born approximation and find that its dependence on $\Lambda$ is similar to the \emph{discrepancy} between the exact solution and that of the   quantum master equation in the Born approximation. We quantify the regime of validity of the quantum master equation in the Born approximation with or without the Markov approximation  in terms of the system's relaxation rate $\gamma$, its \emph{unrenormalized} natural frequency $\Omega$ and $\Lambda$: $\gamma/\Omega \ll 1$ and \emph{also} $\gamma \Lambda/\Omega^2 \ll 1$. The reliability of the Born approximation is discussed within the context of recent experimental settings and more general environments.

\end{abstract}

%\pacs{95.35.+d;12.60.Cn;95.30.Cq}

\maketitle

\section{Introduction, motivation and goals}\label{sec:intro}

The interaction between a quantum mechanical system and a large number of environmental degrees of freedom typically leads to dissipative dynamics of the system's degrees of freedom along with decoherence. The quantum  system plus   environment form a closed dynamical   system described by a density matrix that evolves with unitary time evolution. Tracing out the environmental degrees of freedom lead to the description of the system's dynamics as an \emph{open quantum system}\cite{breuer,weiss,gardiner,ines,daley}. The theory of open quantum systems plays a fundamental role in quantum information and quantum computing since the  quantum mechanical degrees of freedom exploited to implement quantum computing protocols are unavoidably coupled to environmental degrees of freedom. Experimental progress in quantum optics, cavity electrodynamics\cite{braun,girvin} and optomechanics\cite{opto1,opto2,opto3,opto4,kipp} provide novel platforms and architectures for quantum information and computing. The possibility to engineer the properties of the environmental bath\cite{bathengi,ver,paavo,brum,brunner,xu,poya} opens a new, experimentally accessible window into the fundamental aspects of open quantum systems. A recent experiment\cite{groex} has probed the spectral properties of the environmental bath with an opto-mechanical resonator, and experimental studies\cite{exp1,exp2,exp3} have demonstrated the feasibility of coupling various quantum systems to non-equilibrium baths. The experimental possibility to  engineer  the couplings and the environmental degrees of freedom   provides a pathway to controlling the dissipative and decoherence properties of the bath paving the way for experimental control of entanglement of the system's degrees of freedom and could lead to novel cooling techniques of optomechanical systems\cite{xu}.

Theoretical and experimental developments in quantum open systems are also fuelling the emerging field of quantum thermodynamics\cite{quthermo1,quthermo2,quthermo3,paz,buttner1,kosloff,deffner}.

 An important, paradigmatic model of quantum open systems is that of quantum brownian motion\cite{feyver,leggett,grabert,ford1,hu1,hu2osc,flehu,breuer,weiss,carlesso,marzo,feria} that considers the system to be an harmonic oscillator with linear coupling to a bath of a large number of  harmonic oscillators with a continuum density of states.   The properties of this bath are determined by its spectral density. This  simple model has yielded a deep understanding of the role of an  environmental bath on decoherence and dissipation of quantum mechanical degrees of freedom. Not only does this model provide a window into the fundamental questions of decoherence and dissipation, but has also been of fundamental experimental relevance in the design of nanomechanical resonators that work at the quantum limit\cite{haye,blen}.

  A powerful tool to study the non-equilibrium dynamics in quantum open systems is the quantum master equation\cite{gardiner,breuer,ines}. Typically for generic quantum open systems the derivation of the quantum master equation relies on various approximations. Chief  among them are: i) the Born approximation, which assumes a \emph{factorization} between the time dependent reduced density matrix and the time independent density matrix of the environment, usually taken to be in thermal equilibrium, and ii) the Markov approximation which neglects the memory of the time evolution and yields a time-local master equation. Often further approximations, such as the secular or rotating wave approximation are invoked to yield a Lindblad quantum master equation\cite{breuer,gardiner,ines}. The model of quantum Brownian motion provides a case in which the \emph{exact} quantum master equation is available, obtained via path integral techniques\cite{leggett,hu1,ford2,karr,unruh,kumar}, as well as   a similar equation for its Wigner transform\cite{yu}. The exact quantum master equation for quantum Brownian motion is found to be time-local, but is given in terms of time dependent coefficient functions that must be obtained from complicated differential equations with memory kernels\cite{hu1,ford2}. Besides the model of quantum Brownian motion, there are a few other instances in which an exact quantum master equation is available: the case of a (fermionic or bosonic) system of N-independent single particle energy levels linearly coupled to a reservoir of harmonic bosonic modes\cite{nori}, a two level-system (qubit) coupled to a bosonic (harmonic) reservoir\cite{vacchini} (this latter model is exactly solvable because of the total excitation number is conserved). However, more generally such exact solution is not available and the Born-quantum master equation is the tool of choice for studying the non-equilibrium dynamics of quantum open systems. The Born-Markov approximation to the quantum master equation often has shortcomings, for example violations of positivity of the density matrix\cite{breuer,sch}, and further approximations leading to a Lindblad form are often invoked to overcome this problem. More recently a Lindblad model of quantum Brownian motion has been proposed\cite{maciej,kumar} that introduces a time-local Lindblad form for the quantum master equation for quantum Brownian motion that bypasses the complicated time dependent form of the exact master equation and overcomes the shortcomings of the Born-Markov approximations.

 \vspace{2mm}

 \textbf{Motivation.}

 The quantum master equation in the Born approximation, often in Lindblad form\cite{breuer,ines,gardiner}, is widely used in theoretical or experimental studies of non-equilibrium dynamics of open quantum systems. The Born approximation assumes a factorization between the (reduced) density matrix of the system and that of the bath, which is assumed to retain its initial value. This approximation neglects the build up of correlations between the system and the bath  and is typically justified for weak coupling, which under most circumstances implies that the relaxation or damping rate(s) are much smaller than the typical frequency of the system's degrees of freedom. However,   for a given model of the environmental degrees of freedom there are generally several other parameters that characterize the spectral density of the bath, for example the bandwidth (cutoff). The precise manner in which these features of the spectral density determine the regime of reliability of the Born  and Markov  approximations is   seldom discussed either theoretically or experimentally.

  Motivated by the timely importance of quantum master equations to study the dynamics of quantum open systems and by its relevance to the interpretation of experimental results, in this article we focus on establishing a \emph{direct} comparison between the   results for correlation functions obtained with the quantum master equation in the Born approximation with the \emph{exact} results for the case of quantum Brownian motion. To the best of our knowledge there has not yet been a comparison between the exact dynamics and the predictions of the Born quantum master equation (with or without the Markov approximation) for systems with continuous variables.

 \vspace{2mm}

 \textbf{Goals:} Our main goal is to understand the regime of validity of the quantum master equation in the Born approximation by directly comparing its predictions with exact results available in the case of quantum Brownian motion. Whereas the exact quantum master equation for this case has been derived previously and discussed extensively\cite{leggett,hu1,ford2,karr,kumar,yu}, obtaining correlation functions requires the exact solution of this equation. This is in general  a daunting task because the time-dependent coefficients of the quantum master equations must be obtained from the solutions of   differential equations non-local in time. Furthermore, even when this quantum master equations is solved, correlation functions at different times involve the implementation of the quantum regression theorem\cite{breuer,gardiner}.

We proceed instead in a very different manner: we obtain the exact solution of the Heisenberg-Langevin equations which allow us to \emph{directly} obtain any correlation function \emph{exactly} at different times and at all temperatures in terms of the spectral density of the bath. We then focus on the case of an Ohmic bath with a Drude spectral density\cite{breuer,weiss} which offers the distinct advantage  of an exact analytic solution. Such exact analytic treatment allows us to explicitly obtain the dependence of the correlation functions on the various parameters: the bandwidth (cutoff) of the bath $\Lambda$, the system's relaxation rate $\gamma$ and the \emph{unrenormalized} system's natural frequency $\Omega$. A direct comparison with the correlation functions obtained from the quantum master equation in the Born approximation, with or without the Markov approximation, leads to establishing unambiguosly the regime of validity of both approximations.

Furthermore, we study the system-bath correlations to assess the correlations missed by the Born approximation. As a \emph{proxy} for  these correlations we study the interaction energy between the system and the bath, which is also interpreted as the correlation between the system degrees of freedom and the collective bath variable to which the system couples directly.

\section{The model and the Heisenberg-Langevin equation of motion. }\label{sec:model}
We consider an oscillator of unit mass coupled to a bath of oscillators in equilibrium at temperature $T$. The total Hamiltonian is
\be H= H_S+ H_B+ H_{SB} \label{totalH}\ee
\be H_S = \frac{p^2}{2}+ \frac{\Omega^2}{2}~q^2  \,, \label{Hsis}\ee
\be H_B =   \sum_{k} \frac{1}{2} \Big[ {P^2_{k}} + {W^2_{k}} \, Q^2_{k}   \Big]\,. \label{Hbath}\ee with $H_B$  describing a bath of harmonic oscillators to be taken in thermal equilibrium at temperature  $T$. The system-bath coupling is taken to be

\be  H_{SB}   =   -q \, \sum_{k} C_k \,Q_{k} \,,\label{SBcoup} \ee and the  equations of motion   are
  \bea
  && \ddot{q}  + \Omega^2  q     =     \,\sum_{k} C_k \,Q_{k}  \label{EOMq} \\
 && \ddot{Q}_{k} + W^2_{k} Q_{k}   =    C_k\,  q(t)  \,. \label{EQMB}  \eea

 We will solve the  dynamics as an initial condition problem, setting up initial conditions at time $t=0$. This allows us to understand the transient evolution of correlation functions and the approach to an asymptotic stationary state.

We proceed by solving the equations of motion for the bath variables and inserting the solution into the equations of motion for the system coordinate, namely
\bea  Q_{k}(t) & = &  Q^{(0)}_{k}(t) + C_k \int_0^t \frac{\sin\big[W_{k}(t-t') \big]}{W_{k}} \,q(t') \,dt' \,,\label{solB} \eea where
 \be  Q^{(0)}_{k}(t)   =  \frac{1}{\sqrt{2\,W_{k}}}\,\Big[a_k\,e^{-iW_{k}t}+a^\dagger_k\,e^{ iW_{k}t}\Big] \label{zeroQB}  \ee  is the     operator solution  of the homogeneous equation in terms of independent  annihilation ($a_k$) and creation ($a^\dagger_k$) bath operators.

 The   bath is assumed to be in thermal  equilibrium at temperatures $T$ , with statistical average (hereafter we set $\hbar=1~~;~~k_B=1$)
 \be \langle \langle a^\dagger_k a_k \rangle \rangle = \frac{1}{e^{ W_{k}/T}-1} \,,\label{thermal}\ee  where $\langle \langle(\cdots) \rangle \rangle$ define the statistical averages over the bath variables.

  Inserting the solutions (\ref{solB}) into (\ref{EOMq}) we find a   Heisenberg-Langevin equation of motion for the system coordinate,

 \be
\ddot{q}(t) + \Omega^2 q(t) \, +\int_0^t  \Sigma(t-t')\,q(t') dt'  =  \xi(t) \,,\label{HLq} \ee where the self-energy kernel is  given  by

 \be  \Sigma(t-t')  =  -\sum_k \frac{C^2_k}{W_k}\,\sin[W_k(t-t')] \equiv  \frac{i}{\pi}\,\int_{-\infty}^{\infty}  \sigma(\omega') \,e^{i\omega'(t-t')}\,d\omega' \, \label{sigma}\ee   and the spectral density of the bath is
  \be  \sigma(\omega')   = \sum_k \frac{\pi\,C^2_k}{2\,W_{k}}\,\Big[\delta(\omega'-W_{k})-\delta(\omega'+W_{k}) \Big]\,, \label{spec} \ee Note also that $\sigma(\omega) = -\sigma(-\omega)$, and that the noise term is
  \be \xi(t) = \sum_k C_k \, Q^{(0)}_{k}(t) \,.\label{noise} \ee

  A homogeneous version of the Heisenbeg-Langevin equation (\ref{HLq}) (without the noise term and with the lower limit in the non-local integral taken to $-\infty$)  has been considered in refs.\cite{ris,ull} to study dynamical susceptibilities.

  The Heisenberg-Langevin equation of motion (\ref{HLq}) is solved by Laplace transform, for which one introduces

  \be \tilde{q}(s) = \int_0^\infty e^{-st}\,q(t)\,dt~~;~~ \widetilde{\Sigma}(s) =\int_0^\infty e^{-st}\,{\Sigma}(t)\,dt ~~;~~\tilde{\xi}(s) = \int_0^\infty e^{-st}\,{\xi}(t)\,dt \,. \label{laplavars}
  \ee
 From the spectral representation of the self-energy (\ref{sigma}) we find
  \be \widetilde{\Sigma} (s) = - \frac{1}{\pi}\int_{-\infty}^{\infty} \, \frac{\sigma(\omega')}{\omega'+ i\,s}\, {d\omega'}  \,, \label{sigmaofs} \ee and the solution in Laplace is
  \be \tilde{q}(s) = g(s)\Big[ \dot{q}(0) + s\,q(0) +\tilde{\xi}(s)  \Big]\,, \label{sollapla} \ee  where
  \be g(s) = \frac{1}{\big[s^2+ \Omega^2 +\widetilde{\Sigma}(s)\big] }\,.  \label{gofs}\ee
  The solution of Eq. (\ref{HLq}) can be written
  \be q(t) = G(t)\,\dot{q}(0) + \dot{G}(t)\,q(0) + \int_0^t G(t-t') \xi(t')\,dt' \,,\label{qoft}\ee where
  \be G(t) = \int_{\mathcal{C}} \frac{ds}{2\pi i}\, g(s)\, e^{st}  \label{goft} \ee is the solution of (\ref{HLq}) for $\xi(t) =0$    with initial conditions $G(0)=0, \dot{G}(0) =1$, and $\mathcal{C}$ stands for the Bromwich contour, parallel to the imaginary axis and  to the right of all the singularities of the function $g(s)$ in the complex $s$-plane.

   We write the  exact solution (\ref{qoft}) as
  \be q(t) = q_0(t) + q_\xi(t) ~~;~~ q_0(t) = G(t)\,\dot{q}(0) + \dot{G}(t)\,q(0) ~~;~~ q_\xi(t) = \int_0^t G(t-t') \xi(t')\,dt' \label{qsplit} \ee to highlight the   two different contributions: from the initial value ($q_0(t)$) and the noise term ($q_\xi(t)$).

   In the Heisenberg picture the operators depend on time but states do not. Correlation functions of operators are obtained by taking expectation values of the time evolved, Heisenberg operators in the \emph{initial total density matrix} $\rho_{SB}(0)$. For example, the total average
   \be \langle q(t) q(t') \rangle \equiv  \mathrm{Tr}_{SB}~ q(t) q(t') \rho_{SB}(0) \,.\label{qtqtp}\ee

  The exact solution to the Heisenberg-Langevin equation allows to obtain the correlation functions for any \emph{arbitrary initial state}. However,  our goal is to compare to the correlations obtained from the quantum master equation which is usually obtained under the assumption of an initially factorized state. Therefore    we take the \emph{initial} density matrix for system plus bath to be of the factorized form

  \be \rho_{SB}(0) = \rho_S(0)\otimes \rho_B(0) ~~;~~ \rho_B(0) = \frac{e^{-{H_B}/{T}}}{\mathrm{Tr}_B~e^{-{H_B}/{T}} } \,. \label{rhototini}\ee This is the usual choice for the initial density matrix in the quantum master equation approach\cite{breuer,gardiner}, furthermore,  in the Born approximation, such factorization is assumed to hold \emph{at all times}.   While, as discussed in ref.\cite{anglin} such factorization may not yield a correct description of decoherence, we emphasize that our main goal is to compare the exact results from the Heisenberg-Langevin approach to those obtained from the quantum master equation obtained under the usual assumption of factorization.

  Therefore, correlation functions of the system's Heisenberg operators require \emph{two} different averages:

\textbf{(i)} {   Average over the initial conditions, denoted    by $\langle q(t) q(t') \cdots \rangle_S$, correspond to averaging with  the initial   density matrix of the system $\rho_S(0)$. We define averages with $\rho_S(0)$ as \be \langle \,   \mathcal{O} \, \rangle_S \equiv \mathrm{Tr} \rho_S(0) \,\mathcal{O} \,.\label{traceS0def}\ee }

\textbf{(ii)} {   Averages over the noise terms $\xi $ correspond to   averaging over the bath variables with the  statistical distribution of the bath and we define
\be  \langle \langle \,(\cdots) \, \rangle \rangle \equiv  \mathrm{Tr}\rho_B(0) (\cdots)\,.  \label{traceBdef}\ee With $\xi(t)$ given by eqn. (\ref{noise}) and  using (\ref{zeroQB},\ref{thermal}) and the fact that the spectral density satisfies $\sigma(\omega) = -\sigma(-\omega)$, we find

    \be   \langle \langle \xi(t)  \rangle \rangle   =   0   \label{xiexp}\ee    and noise correlator
    \be \langle \langle \xi(t_1) \xi(t_2) \rangle \rangle   =    \frac{1}{\pi} \int^\infty_{-\infty} d\omega ~\sigma(\omega)~n(\omega) ~e^{i\omega(t_1-t_2)} ~~;~~ n(\omega) = \frac{1}{e^{\omega/T}-1} \,. \label{noiseave}\ee  }

    Because the theory is Gaussian, only the one and two- point correlation functions must be obtained; higher order correlation functions can be obtained from Wick's theorem.

 These definitions and  (\ref{xiexp})  yield
    \bea \langle q(t) q(t') \rangle  &= &  G(t) G(t') \langle p^2(0) \rangle_S + \dot{G}(t)\dot{G}(t') \langle q^2(0)\rangle_S + G(t) \dot{G}(t') \langle p(0) q(0) \rangle_S + G(t')\dot{G}(t) \langle q(0) p(0) \rangle_S \nonumber \\ &+& \int^t_0 dt_1 \int^{t'}_0 dt_2 G(t-t_1)\, G(t'-t_2) \,\langle \langle \xi(t_1) \xi(t_2) \rangle \rangle\,, \label{qtqtp2} \eea  where where $p(0) = \dot{q}(0)$. Other correlation functions are obtained in a similar manner.  These results are \emph{exact  and general}; it remains to specify the spectral density of the bath.   We consider an Ohmic bath with a Drude spectral density,
  namely,
  \be \sigma(\omega) = \gamma \omega \,\frac{\Lambda^2}{\Lambda^2+\omega^2} ~~;~~ \gamma >0 \,, \label{ohm} \ee where $\Lambda$ determines the bandwidth of the bath (cutoff). We will consider $\Lambda \gg \Omega, \gamma$ without condition on the ratio $\Lambda/T$ in order to study the cases $\Lambda \gg T,\Lambda \ll T$ in detail. For this spectral density, the symmetrized noise correlation function is given by (\ref{symecor}) in Appendix  \ref{app:integrals}  and  the self-energy (\ref{sigma})  becomes
  \be \Sigma(\tau) = -\gamma \Lambda^2\,e^{-\Lambda |\tau|}\,\mathrm{sign}(\tau)\,.  \label{selfie}\ee  We note that
  \be \Sigma(\tau)~ {}_{\overrightarrow{\Lambda \rightarrow \infty}}~\gamma \frac{d}{d\tau}\,\delta(\tau)\,, \label{limi}\ee
  and its Laplace transform is given by
  \be \widetilde{\Sigma}(s) = -\frac{\gamma \Lambda^2}{\Lambda + s}\,. \label{laplasigma}\ee

  Taking the large $\Lambda$ limit $g(s)$ in (\ref{gofs}) simplifies to
    \be g(s) = \frac{1}{s^2+ \Omega^2_R + \gamma s} \,, \label{gslarLam}\ee where the renormalized frequency is
    \be \Omega^2_R = \Omega^2 - \gamma\Lambda \,.\label{Oren}\ee In this limit we find
    \be G(t) = e^{-\gamma t/2}\,\frac{\sin{Wt}}{W} ~~;~~ W = \sqrt{\Omega^2_R-\frac{\gamma^2}{4}} \,. \label{goftlarLam}\ee In Appendix  \ref{app:GF}  we analyze $g(s)$ and $G(t)$ including the full expression (\ref{laplasigma}) for $\widetilde{\Sigma}(s)$ and confirm the validity of (\ref{goftlarLam}) for $ W/\Lambda \ll 1,\gamma/\Lambda \ll 1$.

    At this point we note that the renormalization condition (\ref{Oren}) restricts the possible values of $\gamma \Lambda$. Since $\Omega^2 > 0$, (as physically expected from an uncoupled oscillator), \emph{if} $\gamma \Lambda > \Omega^2$ it follows that $\Omega^2_R <0$, implying an \emph{instability} due to the fact that  in this case $W$ is purely imaginary, leading to one growing and one decaying mode after coupling to the bath. The existence of a growing solution precludes an asymptotic stationary regime; this can be seen clearly in the correlation function (\ref{qtqtp2}) which would grow without bound as a function of time if $\Omega^2_R <0$.  Therefore, an asymptotically well defined equilibrium or stationary state requires that $\Omega^2_R > 0$, leading to the \emph{stability condition}
    \be \frac{\gamma \Lambda}{\Omega^2} <1  \,. \label{const}\ee

     For $t, t' \gg 1/\gamma$ only the last term of (\ref{qtqtp2}) survives, and a straightforward calculation in this limit yields
     \be  \langle q(t) q(t') \rangle = \frac{1}{\pi} \int^\infty_{-\infty} d\omega ~\frac{\sigma(\omega)~n(\omega)}{\Big[(\omega^2-\Omega^2_R)^2+(\omega\gamma)^2\Big]} ~e^{i\omega(t-t')}\,. \label{asyqtqtp}\ee This correlation function indicates the emergence of a stationary limit for $t,t' \gg 1/\gamma$. With $p(t) = dq(t)/dt$, it is straightforward to obtain in this long time limit
     \be  \langle p(t) p(t') \rangle = \frac{1}{\pi} \int^\infty_{-\infty} d\omega ~\frac{\omega^2\,\sigma(\omega)~n(\omega)}{\Big[(\omega^2-\Omega^2_R)^2+(\omega\gamma)^2\Big]} ~e^{i\omega(t-t')}\,, \label{asyptptp}\ee  and
     \be \langle  (p(t)q(t)+q(t)p(t)) \rangle = \Big[ \frac{d}{dt} + \frac{d}{dt'} \Big]\,\langle q(t) q(t') \rangle\Big|_{t=t'} = 0 \,. \label{asypqpqexa}\ee
     We now focus on the coincidence limit $t=t' \gg 1/\gamma$ to compare to the equal time correlation functions obtained with the quantum master equation in the next section.

     Because $\sigma(\omega)=-\sigma(-\omega)$, it follows that for $t=t'\gg 1/\gamma$
      \bea  \langle \, q^2(t) \, \rangle & = &  \frac{1}{2\pi} \int^\infty_{-\infty} d\omega ~\frac{\sigma(\omega)~\coth[\frac{\omega}{2T}]}{\Big[(\omega^2-\Omega^2_R)^2+(\omega\gamma)^2\Big]} \label{asyqtqtpnew}\\ \langle \, p^2(t) \, \rangle & = &  \frac{1}{2\pi} \int^\infty_{-\infty} d\omega ~\frac{\omega^2\,\sigma(\omega)~\coth[\frac{\omega}{2T}]}{\Big[(\omega^2-\Omega^2_R)^2+(\omega\gamma)^2\Big]} \,. \label{asyptptpnew} \eea Similar exact expressions have been obtained in ref.
      \cite{grabert2}.

      With the Drude spectral density (\ref{ohm}) the integrals can be performed by contour integration, and taking the limits $\Lambda \gg \Omega_R, \gamma$   yields (see appendix (\ref{app:integrals}) for details)
      \be  \langle \, q^2  \, \rangle = \frac{1}{4W} \Bigg\{\coth\Big[ \frac{W+i \gamma/2}{2T}\Big] +\coth\Big[ \frac{W-i \gamma/2}{2T}\Big] \Bigg\} + \frac{\gamma}{2\Lambda^2} ~ i\coth[\frac{i\Lambda}{2T}] + \frac{\gamma}{T^2}F[\Lambda/T, \Omega_R/T,\gamma/T]\,. \label{q2fini}\ee The first term (in brackets) is the contribution from the resonant complex poles at $\omega = i\gamma/2 \pm W$ in the limit $\Lambda \gg W, \gamma$. The second term arises from the pole at $\omega = i\Lambda$ and can be safely neglected for $\Lambda \gg \gamma$. The third contribution arises from the Matsubara poles (from $\coth[\omega/2T]$) at $\omega = 2\pi i l T~~;~~ l=1,2,\cdots$ (see (\ref{integ})); the function F remains finite at all temperatures and $\Lambda$ and is given by eqn. (\ref{FunFu}) in appendix (B).
      In the weak coupling limit, $\Omega_R \gg \gamma$, the first term in (\ref{q2fini}) yields the result for an harmonic oscillator in thermal equilibrium (quantum equipartition).  In particular, in the high temperature limit $T \gg \Omega_R, \gamma$ and $\Lambda \gg \Omega_R, \gamma$ we obtain
       \be  \langle \, q^2  \, \rangle = \frac{T}{\Omega^2_R} + \cdots \,, \label{classq2fini}\ee where the dots stand for terms that are subleading for $T \gg \Omega_R$ and $\Lambda \gg \Omega_R, \gamma$.

      Implementing the same method we find
      \be \langle \, p^2 \, \rangle =  \frac{W}{4} \Bigg\{\big(1+i \gamma/2W\big)^2~\coth\Big[ \frac{W+i \gamma/2}{2T}\Big] +\big(1-i \gamma/2W\big)^2~\coth\Big[ \frac{W-i \gamma/2}{2T}\Big] \Bigg\} + \gamma \,I(\Lambda/T, \Omega_R/T, \gamma/T)\,, \label{p2fini} \ee where the function $I(\Lambda/T, \Omega_R/T, \gamma/T)$ is given by eqn. (\ref{FunIu}) in appendix B. As a consequence of the extra term $\omega^2$ in the integrand in (\ref{asyptptpnew}),  for $\Lambda \gg T,\Omega_R,\gamma$ the dimensionless function $I$  features a logarithmic dependence with the bandwidth;   in this limit we find\cite{ours}
      \be \gamma\,I[\Lambda/T, \Omega_R/T, \gamma/T] = \frac{\gamma}{\pi} \ln\big[ \frac{\Lambda}{2\pi T} \big] + \cdots \label{loga} \ee where the dots stand for terms that remain finite in the limit $\Lambda \rightarrow \infty$\cite{ours}. In the opposite limit $T \gg \Lambda, \Omega_R, \gamma$ this function remains small of the order of the respective ratios displayed in the argument.  In the high temperature limit  with $T \gg \Lambda,\Omega_R,\gamma$ (\ref{p2fini}) yields
      \be \langle \, p^2 \, \rangle = T + \cdots \,,  \label{p2classfini}\ee while for $\Lambda \gg T\gg \Omega_R, \gamma$ we find
      \be   \langle \, p^2 \, \rangle = T\,\Big(1 + \frac{\gamma}{\pi T} \ln\big[ \frac{\Lambda}{2\pi T} \big]\Big) + \cdots \label{p2classloga}\ee The logarithmic dependence on the bandwidth survives in the zero temperature limit, where it yields a correction to $\langle ~ p^2 ~\rangle$ of the form $\gamma/\pi ~ \ln[\Lambda/W]$\cite{ours}.

      The results obtained above are \emph{exact} under the main assumptions $\Lambda \gg \Omega, \gamma$, with  the constraint $\Omega^2 > \gamma \Lambda$ which is necessary to ensure that $\Omega^2_R >0$. This latter constraint implies that the Green's function $G(t)$ (\ref{goftlarLam}) is damped, which leads to an asymptotic stationary state. The results presented above are \emph{general} for a Drude-Ohmic model without restrictions on the ratio $\gamma/\Omega$. The contributions that are proportional to $\Lambda$ (frequency renormalization) and $\ln[\Lambda]$ for $\Lambda \gg T$, are \emph{not} a consequence of the Drude form of the Ohmic spectral density, but a direct consequence of the large separation of scales when $\Lambda \gg \Omega, \gamma$ (the logarithmic correction emerging for $\Lambda \gg T$). For example, it is straightforward to find that an exponential cutoff
      \be \sigma(\omega) = \gamma \omega \, e^{-|\omega|/\Lambda}\,, \label{expocut}\ee instead of the Drude form (\ref{ohm}) yields a similar frequency renormalization ($\propto \gamma \Lambda$) and logarithmic contribution to $\langle p^2 \rangle$ for $\Lambda \gg T,\Omega_R,\gamma$ although the finite temperature integrals are more cumbersome.

      We emphasize that the above results depend on $\Omega_R$ and not directly on the unrenormalized (bare) frequency $\Omega$. This aspect will become an important distinction with the results from the quantum master equation studied below.

      The logarithmic divergence with $\Lambda$ of $\langle p^2 \rangle$ merits discussion. To understand its origin let us consider the correlation function (\ref{asyptptp}) for $t-t' = \tau > 0$ with   $W\tau \ll 1~,~ \gamma \tau \ll 1$ and $\Lambda \gg T \gg \Omega_R, \gamma$. The frequency integral can be done by residues by closing the contour in the upper half $\omega$-plane. The contribution from (complex) zeroes in the denominator in (\ref{asyptptp}) yield the first term (in brackets) in (\ref{p2fini}). The contribution from the poles at $\omega = i \Lambda$ arising from $\sigma(\omega)$ lead to
      \[-i\frac{\gamma}{2}\coth\Big[\frac{i\Lambda}{2T}\Big]~~e^{-\Lambda \tau}\,,\] which is finite and vanishes rapidly   for $\Lambda \tau \gg 1$. The most important contribution arises from the poles corresponding to the Matsubara frequencies $\omega_l = i 2\pi T l $ with $l = 0,1,2\cdots$. These yield the contribution
      \[ \frac{\gamma}{\pi} \sum_{l=1}^\infty \, \frac{1}{l}\,\Big[\frac{(\Lambda/2\pi T)^2}{(\Lambda/2\pi T)^2 - l^2}\Big]  \, e^{-2\pi l T \tau}\,. \]  For $T \gg \Lambda$ this contribution is   $\propto \gamma (\Lambda / T)^2$ and negligible. However, for $\Lambda \gg T$ it can be very well  approximated  for $\tau \gtrsim 1/\Lambda$ by
      \be  \frac{\gamma}{\pi} \sum_{l=1}^\infty \, \frac{1}{l}\,   \, e^{-2\pi l T \tau} = -\frac{\gamma}{\pi}\ln\Big[ 1- e^{-2\pi T \tau}\Big]~~ {}_{\overrightarrow{\tau \rightarrow 0}} ~~-\frac{\gamma}{\pi}\ln\Big[2\pi T\tau\Big] \,.\label{applogterm}\ee Taking the shortest time scale to be that of the bath $\tau \simeq 1/\Lambda$ yields the logarithmic divergence (\ref{loga}) for $T \ll \Lambda$. Therefore, it is clear that the divergence is a consequence of measuring correlations with a time difference shorter than the typical time scale of the bath, which is determined by the bandwidth $\Lambda$. For $\tau \lesssim 1/\Lambda$ the correlation function probes frequencies of the order of the bandwidth of the bath.  Measuring the correlation on time differences   $\tau \gtrsim 1/\Lambda$ cuts-off the divergence with the bandwidth.  A similar conclusion was reached in Ref.\cite{ours} within a different context and for a strictly Ohmic bath.

      This analysis suggests that the bandwidth dependence of correlation functions will be stronger for super-ohmic spectral densities when $T/\Lambda \ll 1$ (see discussion in section (\ref{sec:nonohm})).

\section{Quantum master equation}\label{sec:qme}

In this section we study the dynamics within the framework of the quantum master equation. For the model of quantum Brownian motion under consideration there is a body of work on the \emph{exact} quantum master equation\cite{leggett,hu1,ford2,karr,kumar,yu}. This exact form is obtained from the path integral representation of the time evolution of the total density matrix after tracing over the bath degrees of freedom, which yields an ``influence action''\cite{feyver} for the system's degrees of freedom. The exact master equation so derived is local in time, but involves complicated functions of  time that are obtained from the solution of differential equations with memory. Obviously the \emph{exact} master equation yields the same information as the \emph{exact} solution of the Heisenberg-Langevin equation. While obtaining  correlation functions at different times from the solution of the Heisenberg-Langevin equation is fairly straightforward, as shown in the previous section,  in the master equation approach   obtaining correlation functions at non-equal times requires implementing the quantum regression theorem\cite{breuer,gardiner}.

Our focus here is to compare the exact results obtained above from the solution to the Heisenberg-Langevin equation to the  quantum master equation  often used for  open quantum systems that relies on the Born-Markov approximation valid for \emph{weak coupling}.

Within the context of the model studied here,   the weak coupling condition is equivalent to the condition
\be \gamma/ \Omega_R \ll 1 \label{wccond}\,,\ee  namely, the damping rate is much smaller than the (renormalized) oscillation frequency. As a result of eqn. (\ref{goftlarLam}) this condition is equivalent to    $W \gg \gamma$. Furthermore, consistently with the previous section,  we also assume $\Lambda \gg \Omega_R, \gamma$.

We obtain the quantum master equation in the interaction picture, wherein the full density matrix $\rho_I(t)$ obeys
\be \dot{\rho}_I(t) = -i\Big[H_I(t),\rho_I(t)\Big] \label{rhoIdot}\ee with
$H_I(t) = H_{SB}(t)$ and
$H_{SB}$ is given by Eq. (\ref{SBcoup}). The time evolution is in the interaction picture of $H_S+H_B$, namely $q(t)$ and $Q_k(t)$ evolve in time with the free Hamiltonian. Therefore
\be H_{SB}(t) = -q(t)\, \xi(t) \label{HSBint}\,,\ee where
\be q(t) = q(0)\,\cos(\Omega t) + \frac{p(0)}{\Omega} \, \sin(\Omega t)~~;~~p(0) = \dot{q}(0)\,, \label{qip}\ee and $\xi(t)$ is given by eqn. (\ref{noise}) with $Q^{(0)}_k(t)$, the free bath coordinates, given by (\ref{zeroQB}).

The formal solution is
\be \rho_I(t) = \rho_{SB}(0) - i\int_0^t dt_1 \Big[H_I(t_1),\rho_I(t_1)\Big] \label{rhoIsol}\ee where $\rho_{SB}(0)$ is given by eqn. (\ref{rhototini}). Inserting this solution back into (\ref{rhoIdot}) yields
\be \dot{\rho}_I(t) = -i\Big[H_I(t),\rho_{SB}(0)\Big] - \int^t_0 dt_1 \Big[H_I(t),\Big[H_I(t_1),\rho_I(t_1)\Big]\Big]\,.  \label{rhodotiter}\ee The reduced density matrix for the system variables is obtained by tracing over the bath variables, namely
\be \rho_r(t) = \mathrm{Tr}_{B}~ \rho_I(t)\,,  \label{rhored}\ee and since $\langle \langle \xi(t) \rangle\rangle =0$, the first term in (\ref{rhodotiter}) does not contribute to the quantum master equation for the reduced density matrix. Therefore
\be \dot{\rho}_r(t) = -\mathrm{Tr}_B \int_0^t dt_1 \Big[H_I(t),\Big[H_I(t_1),\rho_I(t_1)\Big]\Big]\,.  \label{rhodotred}\ee In order to obtain the time evolution of expectation values of operators associated with the system's variables in the interaction picture, one would need to solve equation (\ref{rhodotred}) exactly, usually a very difficult task. Instead one obtains evolution equations for the expectation values,
\be \frac{d}{dt} \langle \, \mathcal{O} \,\rangle (t) = \langle \,\dot{\mathcal{O}} \,\rangle (t) + \mathrm{Tr}\mathcal{O}(t) \dot{\rho}_r(t)\,, \label{dotexp}\ee where
\be  \langle \, \mathcal{O} \,\rangle (t) \equiv  \mathrm{Tr}_S\, \mathcal{O}(t) \,\rho_r(t)  \,, \label{expdef}\ee and $\mathcal{O}(t)$ is an operator associated with the system's variables in the interaction picture. Correlation functions at different times may be obtained from the quantum regression theorem\cite{breuer,gardiner}.

Although eqn. (\ref{rhodotred})    is exact but  difficult to solve,  several approximations are usually invoked:

\vspace{2mm}

\textbf{1:) Born (factorization)}

A factorization
\be \rho_I(t) = \rho_r(t)\otimes \rho_B(0) \,,\label{factor}\ee is employed;
under this approximation  the trace over the bath degrees of freedom can be carried out and yields
\bea \dot{\rho}_r(t) & = &  -\int^t_0\,dt_1\Bigg\{g^>(t-t_1)\Big[ q(t)q(t_1)\rho_r(t_1)-q(t_1)\rho_r(t_1)q(t)\Big] \nonumber \\ & + & g^<(t-t_1)\Big[\rho_r(t_1)q(t_1)q(t)-q(t)\rho_r(t_1)q(t_1)\Big]\Bigg\} \,,\label{rhodotfact}\eea  where
$q(t)$ is given by (\ref{qip}). Furthermore,
in terms of the noise variable, $\xi(t)$,   we find (see eqn. (\ref{noiseave}))
\be g^>(t-t_1) = \mathrm{Tr}_{B}\, \xi(t)\, \xi(t_1)\rho_B(0) = \frac{1}{\pi} \int^\infty_{-\infty} d\omega \sigma(\omega) n(\omega) e^{i\omega(t-t_1)} ~~;~~ n(\omega) = \frac{1}{e^{\omega/T}-1}\,,\label{ggreat} \ee
\be g^<(t-t_1) = \mathrm{Tr}_{B}\, \xi(t_1)\, \xi(t)\,\rho_B(0) = \frac{1}{\pi} \int^\infty_{-\infty} d\omega \,\sigma(\omega)\, (1+n(\omega))\, e^{i\omega(t-t_1)} \,.\label{gless} \ee

We note the following identities
\be g^>(t-t_1) = \frac{1}{2\pi} \int^\infty_{-\infty} d\omega \,\sigma(\omega) \, \coth\big[\frac{\omega}{2T}\Big]\, e^{i\omega(t-t_1)} + \frac{i}{2} \Sigma(t-t_1)\,, \label{gpluside}\ee
\be g^<(t-t_1) = \frac{1}{2\pi} \int^\infty_{-\infty} d\omega \,\sigma(\omega) \, \coth\big[\frac{\omega}{2T}\Big]\, e^{i\omega(t-t_1)} - \frac{i}{2} \Sigma(t-t_1)\,, \label{gminide}\ee where $\Sigma(t-t')$ is the self-energy given by Eq. (\ref{sigma}) and (\ref{selfie}) for the Drude-Ohmic case. The symmetrized noise correlation function is given by   $[g^>(t-t_1)+g^<(t-t_1)]$;  its explicit form for the Drude-Ohmic case  is given by (\ref{symecor}) in Appendix (\ref{app:integrals}).

\textbf{2:) Born-Markov approximation}

In this approximation $\rho_r(t_1) \rightarrow \rho_r(t)$ in (\ref{rhodotfact}), and the upper limit of the integral   is taken to $t\rightarrow \infty$. We will consider the Born-Markov approximation while keeping the upper limit at   finite time $t$.

Usually the rotating wave (or secular) approximation is invoked   to yield the quantum master equation in Lindblad form\cite{breuer}, but we will not pursue this further approximation here. With (\ref{dotexp}) and (\ref{expdef}) we find

\bea  \frac{d}{dt} \langle \, q \, \rangle (t) &  = &  \langle \, p \rangle (t) \label{qdot}\\
\frac{d}{dt} \langle \, p \, \rangle (t) &  = &  -\Omega^2 \, \langle \, q \, \rangle (t) + \mathrm{Tr}_S \, p(t) \,\dot{\rho}_r(t) \,, \label{pdot} \eea where
\be \mathrm{Tr}_S \, p(t) \, \dot{\rho}_r(t) = i \int^t_0 \Big[g^>(t-t_1)-g^<(t-t_1)\Big]\mathrm{Tr}_S \,q(t_1) \,\rho_r(t) \,.\label{traza} \ee In the trace in the last expression we need to write the interaction picture operator $q(t_1)$ in terms of operators at time $t$ to identify the last term as an expectation value at time $t$. This is achieved  by inverting the relation (\ref{qip})  for $q(0),p(0)$     to obtain\footnote{Using the identity (\ref{qt1}) the quantum master equation in the Born-Markov approximation can be written in a manifestly time-local form. However, the non-local form is more compact to calculate correlators.}
\bea q(t_1) & = &  q(t) \cos[\Omega(t-t_1)] - \frac{p(t)}{\Omega} \sin[\Omega(t-t_1)] \label{qt1}\\ p(t_1) & = &  p(t) \cos[\Omega(t-t_1)] + \Omega ~ q(t) \sin[\Omega(t-t_1)] \label{pt1}\eea leading to
 \be \mathrm{Tr}_S \, p(t) \, \dot{\rho}_r(t) = \langle \, q \, \rangle (t)\, \alpha(t) - \langle \, p \, \rangle (t) \, \beta(t) \,, \label{trdotp}\ee with the functions $\alpha(t),\beta(t)$ given by (see appendix (\ref{app:integrals}))
 \be \alpha(t) = \gamma \Lambda^2 \mathrm{Re} \Bigg\{\frac{1-e^{-\Lambda t}\,e^{i\Omega t}}{\Lambda - i\Omega} \Bigg\} \,,\label{alfat} \ee
\be \beta(t) = \frac{\gamma \Lambda^2}{\Omega } \mathrm{Im} \Bigg\{\frac{1-e^{-\Lambda t}\,e^{i\Omega t}}{\Lambda - i\Omega} \Bigg\} \,.\label{betat} \ee Taking another time derivative of (\ref{qdot}), using (\ref{pdot}) along with  the above results, we obtain
\be \frac{d^2}{dt^2} \langle \, q \, \rangle (t) + (\Omega^2 -\alpha(t)) \langle \, q \, \rangle (t) + \beta(t) \, \frac{d}{dt} \langle \, q \, \rangle (t) = 0 \,. \label{eomqoft}\ee
Although this differential equation is difficult to solve for all times, for $\Lambda t \gg 1$ and $\Lambda \gg \Omega$ it follows from (\ref{alfat}, \ref{betat}) that
\bea \alpha(t) &  &  {}_{\overrightarrow{\Lambda t \gg 1}} ~~  \gamma \Lambda \,\frac{\Lambda^2}{\Lambda^2 + \Omega^2} ~~ {}_{\overrightarrow{\Lambda   \gg \Omega}} ~~ \gamma \Lambda \label{alfatasy}\\
 \beta(t) & & {}_{\overrightarrow{\Lambda t \gg 1}} ~~ \gamma   \,\frac{\Lambda^2}{\Lambda^2 + \Omega^2} ~~ {}_{\overrightarrow{\Lambda   \gg \Omega}} ~~\gamma  \,. \label{betatasy} \eea   Therefore in this limit the equation of motion (\ref{eomqoft}) becomes
 \be   \frac{d^2}{dt^2}\, \langle \, q \, \rangle (t) +  \Omega^2_R \,  \langle \, q \, \rangle (t) + \gamma \, \frac{d}{dt} \, \langle \, q \, \rangle (t) = 0 \,. \label{eomqoftasy}\ee It is straightforward to confirm that the solution of this equation for $\Lambda t \gg 1$ is the expectation value of eqn. (\ref{qoft}) with the Green's function (\ref{goftlarLam}) and vanishing noise expectation value. In the quantum master equation approach one must find the solution of the quantum master equation for $\rho_r(t)$ from which the correlation functions are obtained by taking the appropriate averages, and the quantum regression theorem yields correlation functions at different times. Instead we   obtain differential equations for  \emph{equal time} correlation functions, which will be compared to the equal time limit ($t \rightarrow t'$ ) of the \emph{exact} correlation functions (\ref{asyqtqtpnew},\ref{asyptptpnew}) in the asymptotic long time limit $t \gg 1/\gamma$. We  find
 \be \frac{d }{dt} \langle \, q^2 \,\rangle(t) = \langle \, (pq+qp)\,\rangle(t) \,, \label{q2ave}\ee

\be \frac{d}{dt}\big[ \langle \, (pq+qp)\,\rangle(t) \big] = 2 \langle \, p^2\,\rangle(t) -2 (\Omega^2 - \alpha(t)) \langle \, q^2\,\rangle(t) -\beta(t) \langle \, (pq+qp)\,\rangle(t) + f(t) \,, \label{pqqp} \ee

 \be \frac{d }{dt} \langle \, p^2 \,\rangle(t) = - (\Omega^2 - \alpha(t))\langle \, (pq+qp)\,\rangle(t) -2 \beta(t) \langle \, p^2 \,\rangle(t) + h(t)  \,, \label{p2ave}\ee where $\alpha(t),\beta(t)$ are given by (\ref{alfat},\ref{betat}), respectively, and
 \bea f(t) & = &  \int^t_0 d\tau \Big[g^>(\tau)+g^<(\tau) \Big]\frac{\sin(\Omega \tau)}{\Omega} \label{foft} \\ h(t) & = &  \int^t_0 d\tau \Big[g^>(\tau)+g^<(\tau) \Big]{\cos(\Omega \tau)} \,. \label{hoft} \eea The explicit form of the integrals for $f(t),h(t)$ are given in Appendix \ref{app:integrals}.

 Solving the set of coupled differential equations (\ref{q2ave}-\ref{p2ave}) is a very difficult task, however we can extract the asymptotic long time limit under the assumption that such limit describes a stationary state. In this limit the left hand side of these equations can be set to vanish and using the asymptotic limits (\ref{alfatasy},\ref{betatasy}) and the results of Appendix  \ref{app:integrals}  (see equations \ref{finfty},\ref{hinfty} ), we find
 \be \langle \, p^2 \,\rangle(\infty) = \frac{h(\infty)}{2\beta(\infty)} = \frac{\Omega}{2}\,\coth[\frac{\Omega}{2T}] \, \label{pinfty} \ee and
  \be   \langle \, q^2 \,\rangle(\infty) =    \frac{\Omega}{2\Omega^2_R}  \,\coth[\frac{\Omega}{2T}]+ \frac{f(\infty)}{2\Omega^2_R} =   \frac{\Omega}{2\Omega^2_R} \,\coth[\frac{\Omega}{2T}]+ \frac{\gamma}{\Omega^2_R}~\Big\{-\frac{T}{\Lambda} - \frac{1}{\pi}\mathrm{Re}\Big[\Psi\Big(\frac{\Lambda}{2\pi T} \Big)-\Psi\Big(\frac{i\Omega}{2\pi T} \Big) \Big] \Big\}  \,.\label{qinfty} \ee The di-Gamma ($\Psi$) function has the following limits \be {\Psi(z)}_{ ~~\overrightarrow{z\rightarrow 0}}~~ -\frac{1}{z} + \mathcal{O}(1) ~~;~~ {\Psi(z)}_{ ~~ \overrightarrow{z\rightarrow \infty}} ~~   \ln(z)+ \mathcal{O}(1/z) \,,\label{digama}\ee  therefore, in the high temperature classical limit $T \gg \Lambda \gg \Omega, \gamma$ we find
  \bea \langle \, p^2 \,\rangle(\infty) & = & T +\cdots \label{p2class} \\
   \langle \, q^2 \,\rangle(\infty) & = & \frac{T}{ \Omega^2_R} \Big(1+ \frac{\gamma}{\Lambda}\Big) + \cdots \label{q2class}\eea where the dots stand for subleading contributions. Hence, we find that for $\Lambda \gg \gamma$ we recover classical equipartition in the high temperature limit when $T \gg \Lambda \gg \Omega, \gamma$. However, in the limit   $\Lambda \gg T \gg \Omega, \gamma$ we find
   \bea \langle \, p^2 \,\rangle(\infty) & = & T +\cdots \label{p2class2} \\
   \langle \, q^2 \,\rangle(\infty) & = & \frac{T}{\Omega^2_R} \Big(1 - \frac{\gamma}{\pi T} \ln\Big[\frac{\Lambda}{2\pi T}\Big] \Big ) +\cdots \label{q2class2}\eea Compare these results to the exact ones (\ref{classq2fini},\ref{p2classloga}).  The difference is striking: In the results (\ref{p2class2}, \ref{q2class2}) the logarithmic dependence on $\Lambda$ appears in $\langle \, q^2 \,\rangle(\infty)$ instead of $\langle \, p^2 \,\rangle(\infty)$ as in the exact results. Although the discrepancy is perturbatively small $\propto \gamma / T$, it is enhanced by the large logarithm of the bandwidth.

   We find that in the Born-Markov approximation, although the differences from the exact results are suppressed by the perturbative ratio $\gamma/T$, these are \emph{enhanced} by $\ln[\Lambda/T]$, which could be a substantial enhancement for large bandwidth and low  temperature. The logarithmic enhancement appears in $\langle \, q^2 \,\rangle(\infty)$ whereas the exact solution features this enhancement in $\langle \, p^2 \,\rangle(\infty)$ \emph{only}.

   Another important difference emerges when writing the asymptotic correlation functions in terms of the \emph{renormalized} frequency in order to compare to the exact results (\ref{asyqtqtpnew},\ref{asyptptpnew}) in the equal time limit. For example, compare (\ref{pinfty})  and the first term in (\ref{qinfty}) with   the exact expressions (\ref{q2fini},\ref{p2fini}) (neglecting the term proportional to $\gamma$ in both cases); even neglecting terms of order $\gamma^2/\Omega^2_R$ in the arguments of the $\coth[(\cdots)]$ in (\ref{q2fini}), the two expressions differ in the frequency dependence, the exact solution depending \emph{solely} on the renormalized frequency $\Omega_R$ whereas the solution in the Born-Markov approximation depends both on the bare and renormalized frequencies. In weak coupling,  the Born-Markov approximation coincides with the exact result  \emph{only} in the classical limit $T \gg \Omega, \Omega_R , \gamma$.

   \vspace{2mm}

\textbf{3:) Non-Markovian evolution: keeping the memory}
The quantum master equation after the Born (factorization) approximation but \emph{without} the Markov approximation is given by Eq. (\ref{rhodotfact}). We now consider this equation keeping the memory, namely with $\rho_r(t_1)$ in the integral. Using Eq. (\ref{dotexp}) we find
\be \frac{d}{dt} \langle q \rangle (t) = \langle p \rangle (t) \,  \label{dqdtnmk}\ee and
\be \frac{d}{dt} \langle p \rangle (t) = -\Omega^2 \,\langle q \rangle (t) + i\int^t_0    \Big[ g^>(t-t_1) - g^<(t-t_1)\Big] \, \langle q \rangle (t_1) \, dt_1 \,.  \label{dpdtnmk} \ee  Taking another time derivative of (\ref{dqdtnmk}), using equation (\ref{dpdtnmk}) and  the definitions (\ref{ggreat}-\ref{gminide}) we find
\be \frac{d^2}{dt^2}\, \langle q \rangle (t) + \Omega^2 \langle q \rangle (t) + \int^t_0 \Sigma(t-t_1)\,\langle q \rangle (t_1) dt_1 =0 \,.\label{diffeqnmk}\ee This is \emph{precisely} the expectation value of the   Heisenberg-Langevin equation (\ref{HLq}), because the expectation value of the noise term vanishes.

In the non-Markovian master equation (\ref{rhodotfact}) the reduced density matrix appears evaluated at time $t_1$ the expectation values of operators as defined by Eq. (\ref{expdef}) require the operators in the interaction picture to be evaluated at the (integrated) time $t_1$. To obtain the equations of motion for the equal time correlation functions we need the following identities for interaction picture operators
\bea q(t) & = &  q(t_1) \cos[\Omega(t-t_1)] + p(t_1) \frac{\sin[\Omega(t-t_1)]}{\Omega} \label{qiuti}\\p(t) & = &  p(t_1) \cos[\Omega(t-t_1)] - q(t_1) \Omega \sin[\Omega(t-t_1)] \,. \label{piti}   \eea Using these identities it is straightforward to obtain the following hierarchy of coupled equations,
\be \frac{d}{dt} \Big[ \langle q \rangle (t) \Big] = \langle (pq+qp) \rangle (t) \,, \label{dq2dtnmk}\ee
\be \frac{d}{dt} \Big[ \langle (pq+qp) \rangle (t) \Big] = 2 \langle p^2 \rangle (t) - 2 \Omega^2\,\langle q^2 \rangle (t) + \int^t_0 \Big[2 C(t-t_1)\, \langle q^2 \rangle (t_1) + S(t-t_1)\, \langle (pq+qp) \rangle (t_1)  \Big] dt_1 + f(t)\,, \label{dpqdtnmk}\ee
\be \frac{d}{dt} \Big[ \langle p^2 \rangle (t) \Big] = - \Omega^2   \langle (pq+qp) \rangle (t) +\int_0^t \Big[C(t-t_1)\, \langle (pq+qp) \rangle (t_1) -2 \Omega^2\,S(t-t_1)\,\langle q^2 \rangle (t_1)  \Big] dt_1 + h(t) \,,\label{dp2dtnmk}\ee where
\bea C(t-t_1) & = &  i\Big[ g^>(t-t_1)- g^<(t-t_1)\Big]\,\cos[\Omega(t-t_1)] \,,\label{Ctt1}\\
S(t-t_1) & = &  i\Big[ g^>(t-t_1)- g^<(t-t_1)\Big]\,\frac{\sin[\Omega(t-t_1)]}{\Omega} \,,\label{Stt1}
\eea with $f(t)$ and $h(t)$ given by Eqns. (\ref{foft}),\ref{hoft}) respectively.
  The set of integro-differential coupled equations, (\ref{dq2dtnmk},\ref{dpqdtnmk},\ref{dp2dtnmk}) can be turned into an algebraic system of three inhomogeneous coupled equations with three unknowns by Laplace transform. The time evolution of the individual expectation values  is then obtained by performing the inverse transform. If a stationary state emerges asymptotically, the equal time correlation functions become independent of time, therefore in the asymptotic long time limit the inverse Laplace transform of the equal time correlation functions is dominated by an isolated pole at $s=0$\footnote{If a singularity at $s=0$ is not isolated but is the end-point of a branch cut, the correlations will vanish asymptoticaly with long-time tails. The exact solution of section (\ref{sec:model}) shows the emergence of an asymptotic stationary state exponentially.} (Laplace variable).  Obtaining the Laplace transforms and solving the resulting inhomogeneous algebraic system is straightforward. However, finding the poles, and in particular the isolated pole at $s=0$ that yields the asymptotic long time behavior is not. Instead, we   propose a method that systematically yields a derivative expansion and, as argued below, under the assumption that asymptotically there emerges a stationary state, it gives the \emph{exact} asymptotic behavior of the equal time correlators. Consider the following generic term in the memory integrals in the above expressions,
\be I(t) = \int^t_0 K_0(t;t_1) \langle \,\mathcal{O}\,\rangle (t_1) dt_1 \, , \label{Ioft} \ee and write
\be K_0(t;t_1) = \frac{d}{dt_1}\,K_1(t;t_1) ~~;~~ K_1(t;t_1) \equiv \int^{t_1}_0 K_0(t;t_2) \, dt_2 ~~;~~ K_1(t;0) =0 \,.\label{Kone} \ee Then
\be I(t) = K_1(t;t) \,\langle \, \mathcal{O} \, \rangle (t) - \int^t_0 K_1(t;t_1) \frac{d}{dt_1} \Big[  \langle \,\mathcal{O}\,\rangle (t_1) \Big] \,dt_1 \,. \label{I1oft}\ee Repeating the procedure and writing

\be K_1(t;t_1) = \frac{d}{dt_1} \, K_2(t;t_1)    ~~;~~ K_2(t;t_1)     \equiv \int^{t_1}_0 K_1(t;t_2) \, dt_2 ~~;~~ K_2(t;0) =0 \,,\label{Ktwo} \ee the integral in (\ref{I1oft}) is again performed by integration by parts with the result
\be I(t) = K_1(t;t) \, \langle \,\mathcal{O}\,\rangle (t) - K_2(t;t)\, \frac{d}{dt}\Big[\langle \,\mathcal{O}\,\rangle (t) \Big] + \int^t_0 K_2(t;t_1) \frac{d^2}{dt^2_1}\,\Big[  \langle \,\mathcal{O}\,\rangle (t_1) \Big] dt_1 \,.\label{I3oft} \ee Iterating this procedure one finds
\be I(t) = K_1(t;t) \,\langle \, \mathcal{O} \, \rangle (t) -   K_2(t;t) \frac{d}{dt} \Big[  \langle \,\mathcal{O}\,\rangle (t) \Big] +  K_3(t;t) \frac{d^2}{dt^2} \Big[  \langle \,\mathcal{O}\,\rangle (t) \Big] + \cdots\,, \label{Ifinoft}\ee with
\bea K_1(t;t) & = & \int^t_0 K_0(t;t_1) \,dt_1 \label{K1fin} \\
K_2(t;t) & = & \int^t_0 dt_1 \int^{t_1}_0 K_0(t;t_2) \,dt_2 \label{K2fin}\\
K_3(t;t) & = & \int^t_0 dt_1 \int^{t_1}_0 dt_2 \int^{t_2}_0 K_0(t;t_3) \,dt_3 \,. \label{K3fin}\\ \vdots & = & \vdots \eea In order to understand the physical nature of this expansion, consider the kernel
\be K_0(t;t_1) = i \Big[g^>(t-t_1)-g^<(t-t_1)\Big]\,e^{i\Omega(t-t_1)} \, \label{Kexpo}\ee whose real and imaginary parts yield $C(t-t_1)$ and $S(t-t_1)$  in equations (\ref{Ctt1},\ref{Stt1}) respectively, which enter in the hierarchy of equations (\ref{dpqdtnmk},\ref{dp2dtnmk}). With the result (\ref{ggmingl}) in Appendix (\ref{app:integrals}) we find

\be K_1(t;t) = \frac{\gamma \,\Lambda^2}{(\Lambda-i\Omega)}\,\Big[1-e^{-\Lambda t}\,e^{i\Omega t} \Big] \,, \label{K1expo}\ee
\be K_2(t;t) = \frac{\gamma \,\Lambda^2}{2(\Lambda-i\Omega)^2 }\,\Bigg[1-e^{-\Lambda t}\,e^{i\Omega t}\,\Big(1+ (\Lambda -i\Omega)t\Big)\Bigg] \,.\label{K2expo}\ee Therefore, it is clear that for $t \gg 1/\Lambda$, the kernels $K_2(t;t), K_3(t;t) \cdots$ that multiply the derivatives involve higher powers of $1/\Lambda$, namely the time scale of relaxation of the bath degrees of freedom.  Using these results for $C(t-t_1)$ and $ S(t-t_1)$, we now write (\ref{dpqdtnmk},\ref{dp2dtnmk}) in the derivative expansion
\bea \frac{d}{dt} \Big[ \langle (pq+qp) \rangle (t) \Big] & = &  2 \langle p^2 \rangle (t) - 2 (\Omega^2- C_1(t;t)) \,\langle q^2 \rangle (t) + S_1(t;t) \, \langle (pq+qp) \rangle (t)    + f(t) \nonumber \\ & + &
  2\,C_2(t;t)\,\frac{d}{dt} \Big[\langle q^2 \rangle (t)\Big] + S_2(t;t) \,  \frac{d}{dt} \Big[ \langle (pq+qp) \rangle (t) \Big]+\cdots \,, \label{dpqdtnmkder}\eea

\bea \frac{d}{dt} \Big[ \langle p^2 \rangle (t) \Big] & = &  - (\Omega^2-C_1(t;t))   \langle (pq+qp) \rangle (t)    -2 \Omega^2\,S_1(t;t)\,\langle q^2 \rangle (t)    + h(t)\nonumber \\ & + & C_2(t;t)\,  \frac{d}{dt} \Big[ \langle (pq+qp) \rangle (t) \Big]- 2\Omega^2\, S_2(t;t) \,\frac{d}{dt} \Big[\langle q^2 \rangle (t)\Big]+ \cdots  \,,\label{dp2dtnmkder} \\
C_1(t;t) & = &  \alpha(t) ~~;~~ S_1(t;t) = \beta(t) \,,\label{C1S1}\eea where $\alpha(t)$ and $\beta(t)$ are given by (\ref{alfat},\ref{betat}) respectively. For $\Lambda t \gg 1; \Lambda \gg \Omega$ we find
\be C_1(t;t){}_{\overrightarrow{ \Lambda t \gg 1,\Lambda   \gg \Omega}} ~~ \gamma \Lambda ~~;~~ C_2(t;t) ~~ {}_{\overrightarrow{\Lambda t \gg 1,\Lambda   \gg \Omega}}~~\gamma ~~;~~ S_1(t;t) ~~ {}_{\overrightarrow{\Lambda t \gg 1,\Lambda   \gg \Omega}}~~ {\gamma} ~~;~~ S_2(t;t) ~~ {}_{\overrightarrow{\Lambda t \gg 1,\Lambda   \gg \Omega}}~~\frac{\gamma}{\Lambda} \,, \label{2coefs}\ee
 and the dots stand for (higher) derivative terms. Under the assumption that at asymptotically long times the reduced density matrix describes a \emph{stationary state} with equal time correlation functions   independent of time, all derivative terms in the above equations \emph{vanish} asymptotically, and the derivative expansion yields the \emph{exact  relation between the asymptotic values beyond the  Markov approximation}  with     the Born approximation \emph{only}, namely factorization. Therefore,   any discrepancies with the exact results must be identified as being a consequence of the Born approximation.

 The equal time correlation functions in the asymptotic stationary state are obtained by setting  to zero all derivatives in (\ref{dq2dtnmk},\ref{dpqdtnmkder},\ref{dp2dtnmkder}) leading to
 \be \langle (pq+qp) \rangle (\infty) = 0 \,,\label{asypqcorr}\ee
 \be \langle q^2 \rangle (\infty) = \frac{h(\infty)}{2\beta(\infty)\Omega^2} = \frac{1}{2\Omega} \, \coth\Big[ \frac{\Omega}{2T} \Big]\,, \label{asyq2}\ee
 \be  \langle p^2 \rangle (\infty)    =     \Omega^2_R \,\langle q^2 \rangle (\infty)- \frac{f(\infty)}{2} ~  =  ~  \frac{\Omega^2_R}{2\Omega} \, \coth\Big[ \frac{\Omega}{2T} \Big]+ \gamma\,\Bigg\{\frac{T}{\Lambda} + \frac{1}{\pi}\,\mathrm{Re} \Bigg[\Psi\big( \frac{\Lambda}{2\pi T}\big)- \Psi\big( \frac{i\Omega}{2\pi T}\big) \Bigg] \Bigg\}\,. \label{asyp2} \ee In the high temperature classical limit $T \gg \Lambda \gg \Omega_R, \gamma$, we find
 \bea \langle \, p^2 \,\rangle(\infty) & = & T\,\Big[1 - \frac{\gamma \Lambda}{\Omega^2}  +\cdots \Big] \,,\label{p2classnmk} \\
   \langle \, q^2 \,\rangle(\infty) & = & \frac{T}{ \Omega^2}\Big[1   +\cdots \Big] = \frac{T}{ \Omega^2_R}\Big[1 - \frac{\gamma \Lambda}{\Omega^2}  +\cdots \Big]\,, \label{q2classnmk}\eea where   the relation (\ref{Oren}) has been used.  In the limit $\Lambda \gg T \gg \Omega, \gamma$ the result (\ref{asyp2}) becomes
    \be  \langle p^2 \rangle (\infty)    =  T\,\Big[1 - \frac{\gamma \Lambda}{\Omega^2}\, + \frac{\gamma}{\pi T} \ln\Big[\frac{\Lambda}{2\pi T} \Big] +\cdots \Big]\,,  \label{asylam2}\ee
   where the dots stand for subleading contributions, and we have used the relation (\ref{Oren}) between the renormalized and bare frequencies. We emphasize that we are considering the full non-Markovian evolution of the reduced density matrix; the \emph{only} approximation is the Born (or factorization) approximation, but the evolution keeps the memory in the quantum master equation. Comparison of the above results to the exact ones in the high temperature limit, (\ref{classq2fini},\ref{p2classfini}) reveals important differences. We first note that in agreement with the exact result (\ref{p2classfini}), now the logarithimic enhancement is in $\langle p^2 \rangle$; the main discrepancies in the high temperature limit can be seen comparing (\ref{classq2fini}) with (\ref{q2classnmk}) in the high temperature limit and  the first two  terms in the bracket in (\ref{asylam2}) compared to those in Eq. (\ref{p2classloga}). It is clear that the   results  obtained with the exact Heisenberg-Langevin equations agree with those obtained with the Born-non-Markov quantum master equation   when $\gamma/\Omega_R \ll 1$ \emph{and}  $\gamma \Lambda/\Omega^2 \ll 1$. We note also that whereas the exact result (\ref{q2fini}) depends \emph{solely} on the renormalized frequency, the Born-non-Markov result (\ref{asyq2}) depends on the \emph{bare} frequency.  The differences between the exact results and those obtained with the  \emph{Born-non-Markov} quantum master equation can only  originate in the system-bath correlations that are missed by the Born approximation (factorization). The analysis above shows that  these correlations yield corrections of order $\gamma \Lambda /\Omega^2$. These can be substantial when $\Lambda/\Omega \gg 1$ even in  the weak coupling limit $\gamma /\Omega \ll 1$.

   \vspace{2mm}

   \textbf{Comparison: Exact vs. Born-Markov, vs. Born-Non-Markov}

   We now  summarize the comparison between the exact solution and those obtained from the quantum master equation with the Born approximation with and \emph{without} the Markov approximation.

   \textbf{Equation of motion.}

  The   equation of motion for the expectation value $\langle q \rangle (t)$ obtained in the Born-Non-Markov case (\ref{diffeqnmk}) coincides with the expectation value in the initial density matrix of the  \emph{exact} solution of the Heisenberg-Langevin equation (\ref{HLq}) because the average of the noise term vanishes. For the Markovian case the equation of motion (\ref{eomqoft}) differs from the exact and non-Markovian versions at early time, but its solution for $\Lambda t \gg 1$ coincides with that of the exact and non-Markovian cases.

     \textbf{Equal time correlation functions:} Comparing the equal time correlation functions obtained from the quantum master equation in the Born approximation in the asymptotic stationary state we find remarkable differences between the Markov and non-Markov approximation that originate in the \emph{memory} in the non-Markov case. Consider   $\langle q^2 \rangle (\infty)$ and $\langle p^2 \rangle (\infty)$ : The results with the Markov approximation are given by (\ref{qinfty},\ref{pinfty}) respectively, whereas the results obtained by keeping the memory (non-Markovian) are given by (\ref{asyq2},\ref{asyp2}) respectively.

        The difference between the Markovian (M) and the non-Markovian (NM) asymptotic results yield a quantitative measure of the \emph{memory effects}. In the high temperature regime   for $T \gg \Lambda \gg \Omega,\gamma$ we find
        \bea \langle p^2 \rangle_{NM}(\infty)-\langle p^2 \rangle_{M}(\infty) & = & -T\Big[\frac{\gamma \Lambda}{\Omega^2}  + \cdots \Big] \label{p2diff1}\\
         \langle q^2 \rangle_{NM}(\infty)-\langle q^2 \rangle_{M}(\infty) & = & -\frac{T}{\Omega^2_R}\Big[\frac{\gamma \Lambda}{\Omega^2}  + \cdots \Big]\,, \label{q2diff1} \eea while for $\Lambda \gg T \gg \Omega,\gamma$ we find
          \bea \langle p^2 \rangle_{NM}(\infty)-\langle p^2 \rangle_{M}(\infty) & = & -T ~\Big(\frac{\gamma \Lambda}{\Omega^2} \Big)\, \Big[1-\frac{\Omega^2}{\pi T\,\Lambda}\,\ln\Big[\frac{\Lambda}{2\pi T} \Big] + \cdots \Big] \label{p2diff2}\\
         \langle q^2 \rangle_{NM}(\infty)-\langle q^2 \rangle_{M}(\infty) & = & -\frac{T}{\Omega^2_R}~\Big(\frac{\gamma \Lambda}{\Omega^2} \Big)\Big[1-\frac{\Omega^2}{\pi T\,\Lambda}\,\ln\Big[\frac{\Lambda}{2\pi T} \Big] + \cdots \Big]\,. \label{q2diff2} \eea In the expressions above the dots stand for terms $\mathcal{O}(\gamma^2)$. The coefficients of the logarithmic terms are very small in the range considered.

         We conclude that in the high temperature limit the main discrepancies  in the correlation functions  between the non-Markovian and Markovian cases arising from memory effects are: \textbf{i:)}   frequency renormalization since $\gamma \Lambda/\Omega^2 = 1 -\Omega^2_R/\Omega^2 $, \textbf{ii):} the  logarithmic term $\propto \ln[\Lambda/2\pi T]$, although this term is subleading because it  is suppressed by $\gamma/T \ll 1$  compared to the first term.

         The origin of the discrepancies between the Markov (M) and non-Markov (NM) results in the asymptotic long time limit within the Born approximation can be traced to the difference between Eq. (\ref{p2ave}) and the first line in Eq. (\ref{dp2dtnmkder}). In the Markov approximation the density matrix $\rho_r$ is evaluated at time $t$ and the averages are defined with this time argument. Therefore the relations (\ref{qt1},\ref{pt1}) must be used to express the interaction picture operators with argument $t_1$ in terms of those with arguments at time $t$. In the non-Markov case $\rho_r$ is at time $t_1$, therefore the operators at time $t$ must be related to those at time $t_1$ by the relations (\ref{qiuti},\ref{piti}).  This is tantamount to exchanging $q$ and $p$ in the time integrals and also explains the change in sign in the term proportional to $\langle (pq+qp) \rangle$ in Eqs.(\ref{pqqp},\ref{dpqdtnmkder}).

        We have argued above that within the Born approximation in the non-Markov case, the equal time correlation functions in the asymptotic stationary state are \emph{exact}. Therefore any differences with the exact results obtained from the solution of the Heisenberg-Langevin equation are a consequence of the Born approximation.

         Comparing the results of the non-Markov (NM) case (\ref{asyq2},\ref{asyp2}) with the exact (E) results in the high temperature limit (\ref{classq2fini},\ref{p2classfini},\ref{p2classloga}) also shows important discrepancies, in both cases   $T \gg \Lambda \gg \Omega, \gamma$ and  $\Lambda \gg T \gg \Omega, \gamma$
          \bea \langle p^2 \rangle_{E}(\infty) - \langle p^2 \rangle_{NM}(\infty) & = & T \Big[\frac{\gamma \Lambda}{\Omega^2} +\cdots \Big] \label{pdiffENM1} \\
          \langle q^2 \rangle_{E}(\infty) - \langle q^2 \rangle_{NM}(\infty) & = & \frac{T}{\Omega^2_R} \Big[\frac{\gamma \Lambda}{\Omega^2} +\cdots \Big]\,, \label{qdiffENM1} \eea

         \noindent  Therefore the reliability of the Born approximation  in the high temperature limit requires the following \emph{two} conditions
         \be \frac{\gamma}{\Omega}  \ll 1 ~~~ \textit{and}~~~\frac{\gamma \Lambda}{\Omega^2} \ll 1\,. \label{bornconds}\ee  The first is the usual weak coupling condition,  the second is a new condition, it   is equivalent to $|\Omega^2_R-\Omega^2| \ll \Omega^2$ and  is much more restrictive since a  ``coarse grained'' effective description of the system  requires a wide separation of scales, with $\Lambda / \Omega \gg 1$. This analysis shows that the discrepancies between the exact correlations and those obtained in the Born approximation are $\propto \gamma \Lambda/\Omega^2$ and $(\gamma/T)\,\ln\big[\Lambda/2\pi T\big]$, leading to the central conclusion that   the system-bath correlations that are missed in the Born approximation contain these $\Lambda$ dependent terms.

\begin{figure}[!h]
\begin{center}
\includegraphics[height=3.5in,width=3.5in,keepaspectratio=true]{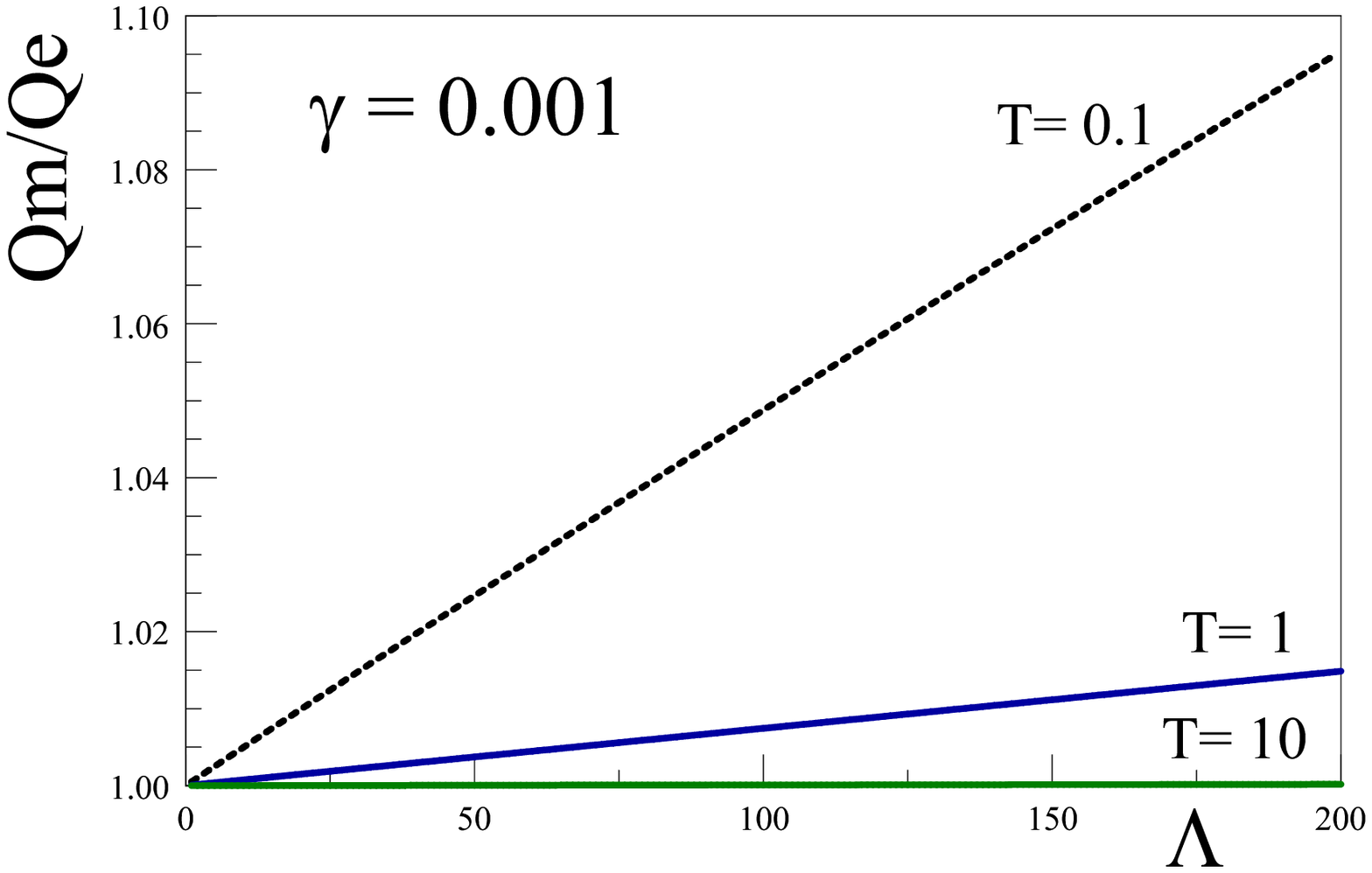}
\includegraphics[height=3.5in,width=3.5in,keepaspectratio=true]{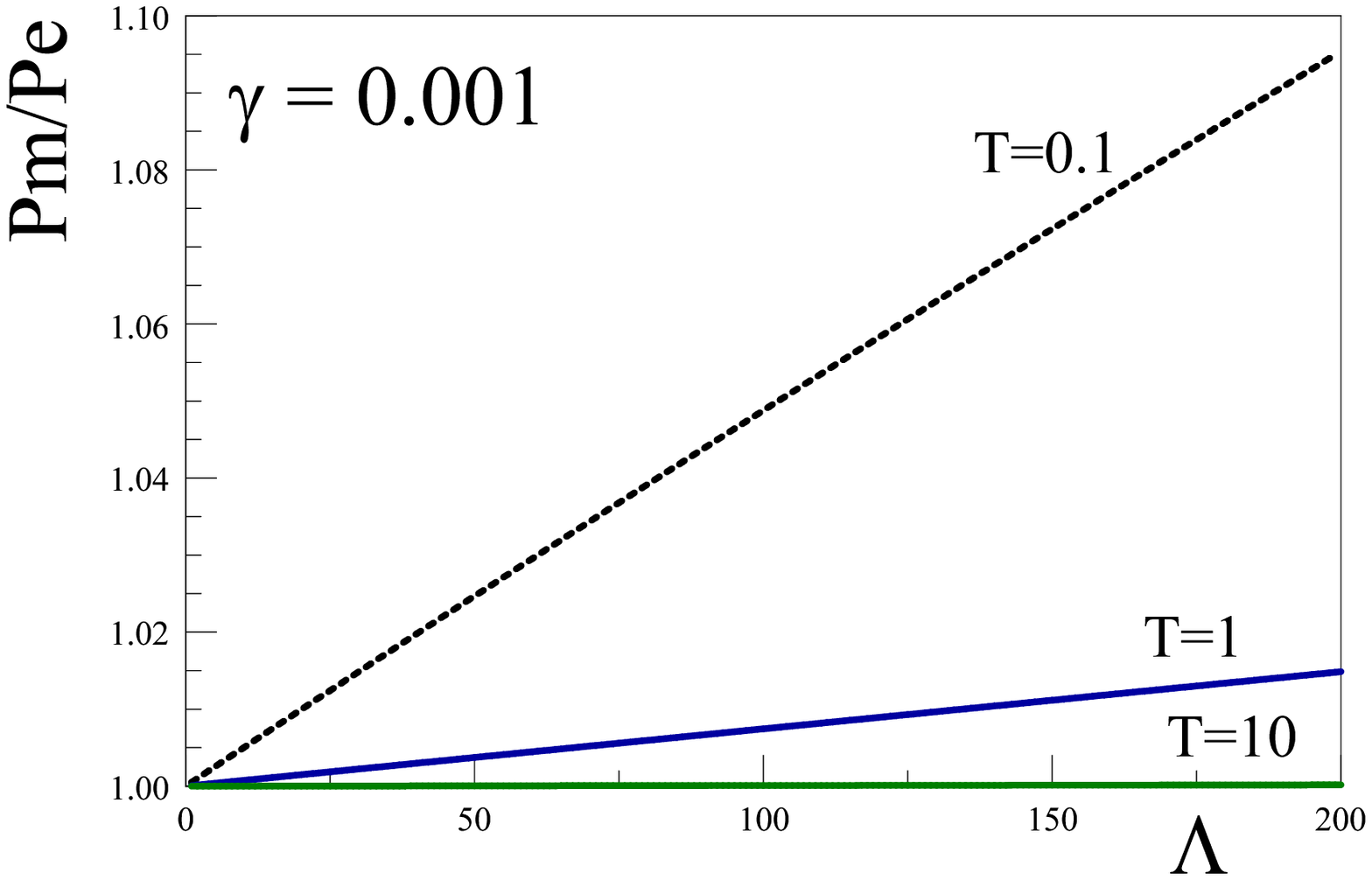}
\includegraphics[height=3.5in,width=3.5in,keepaspectratio=true]{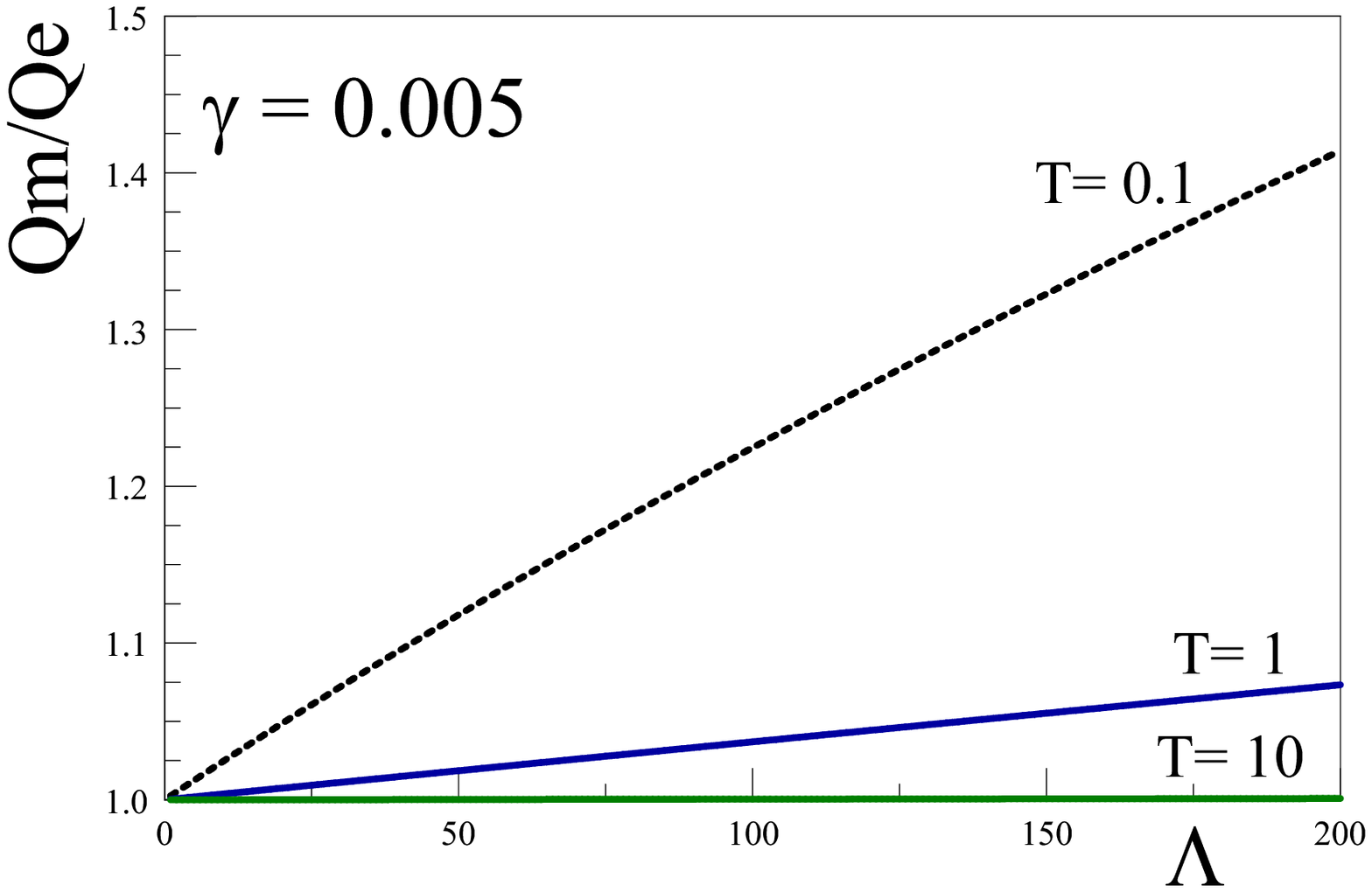}
\includegraphics[height=3.5in,width=3.5in,keepaspectratio=true]{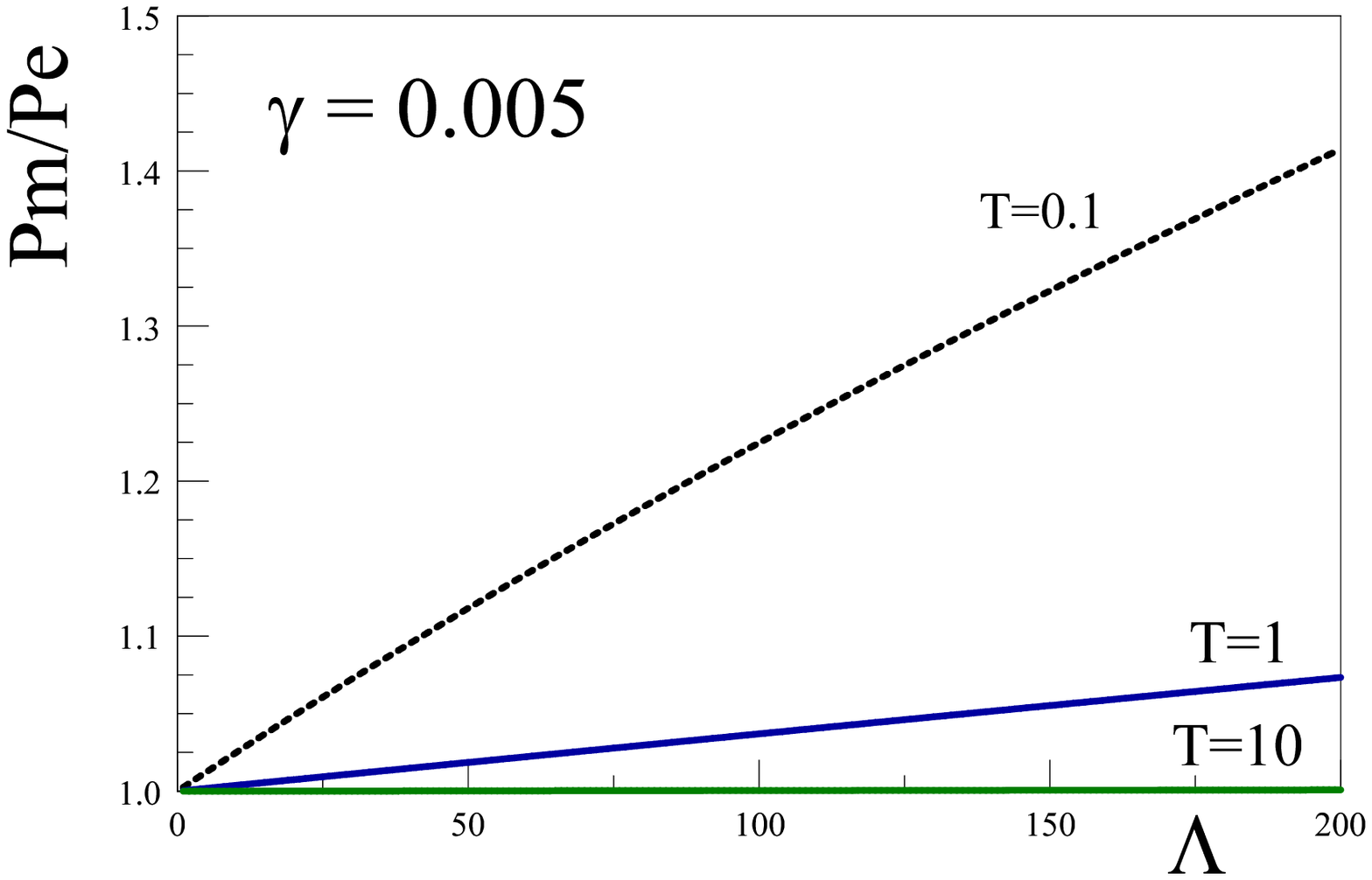}
\caption{ Markovian case: the ratios $\frac{\langle q^2(\infty)\rangle_{M}}{\langle q^2(\infty)\rangle_E} \equiv Qm/Qe$, and $\frac{\langle p^2(\infty)\rangle_M}{\langle p^2(\infty)\rangle_E} \equiv Pm/Pe$, for $\gamma = 0.001~,~0.005$ vs. $\Lambda$ for various temperatures. $\gamma, \Omega, T,\Lambda$ are in units of $\Omega_R$.   At high temperatures the ratios approach $1$. }
\label{fig:markovratios}
\end{center}
\end{figure}

\begin{figure}[!h]
\begin{center}
\includegraphics[height=3.5in,width=3.5in,keepaspectratio=true]{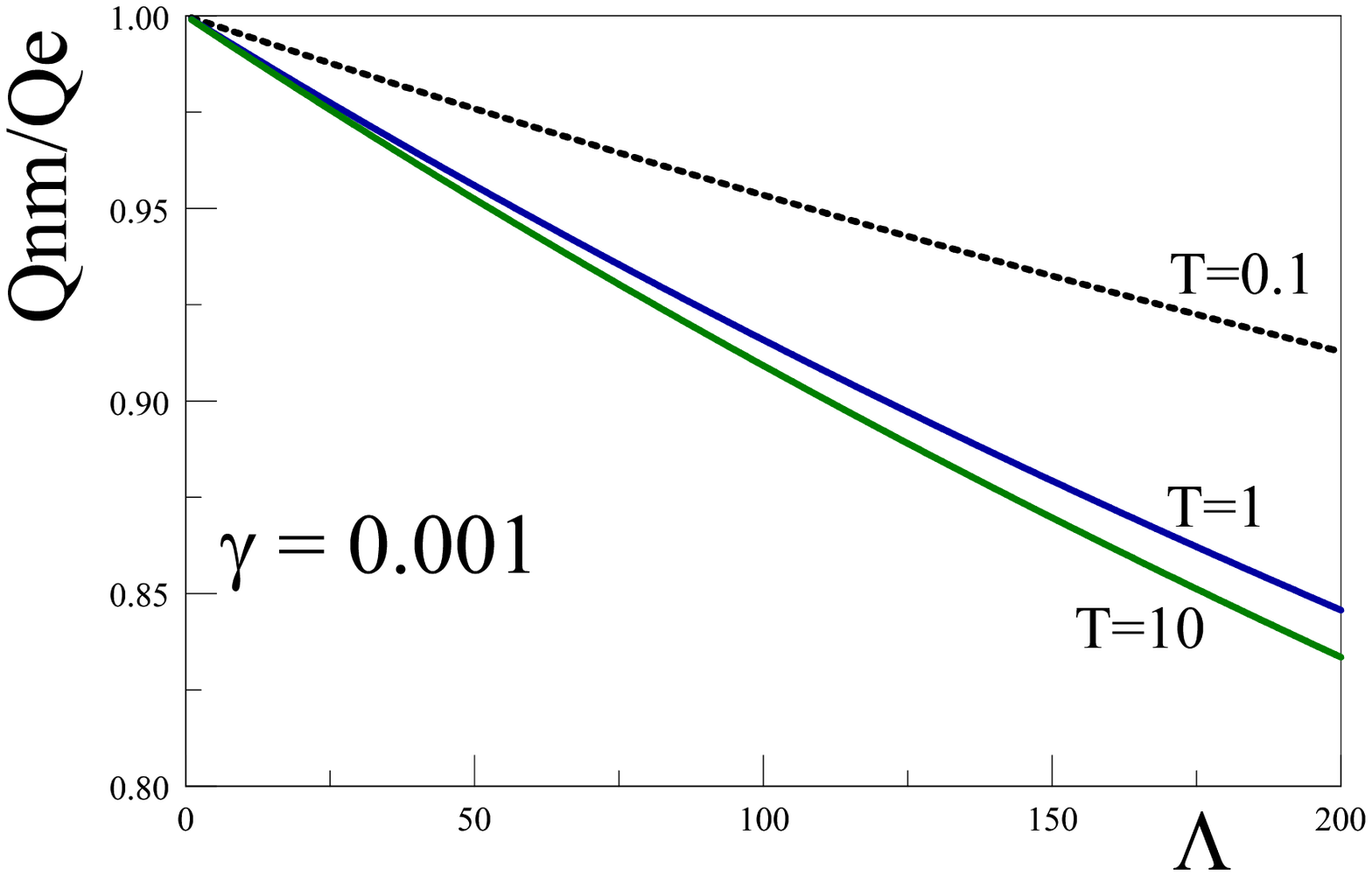}
\includegraphics[height=3.5in,width=3.5in,keepaspectratio=true]{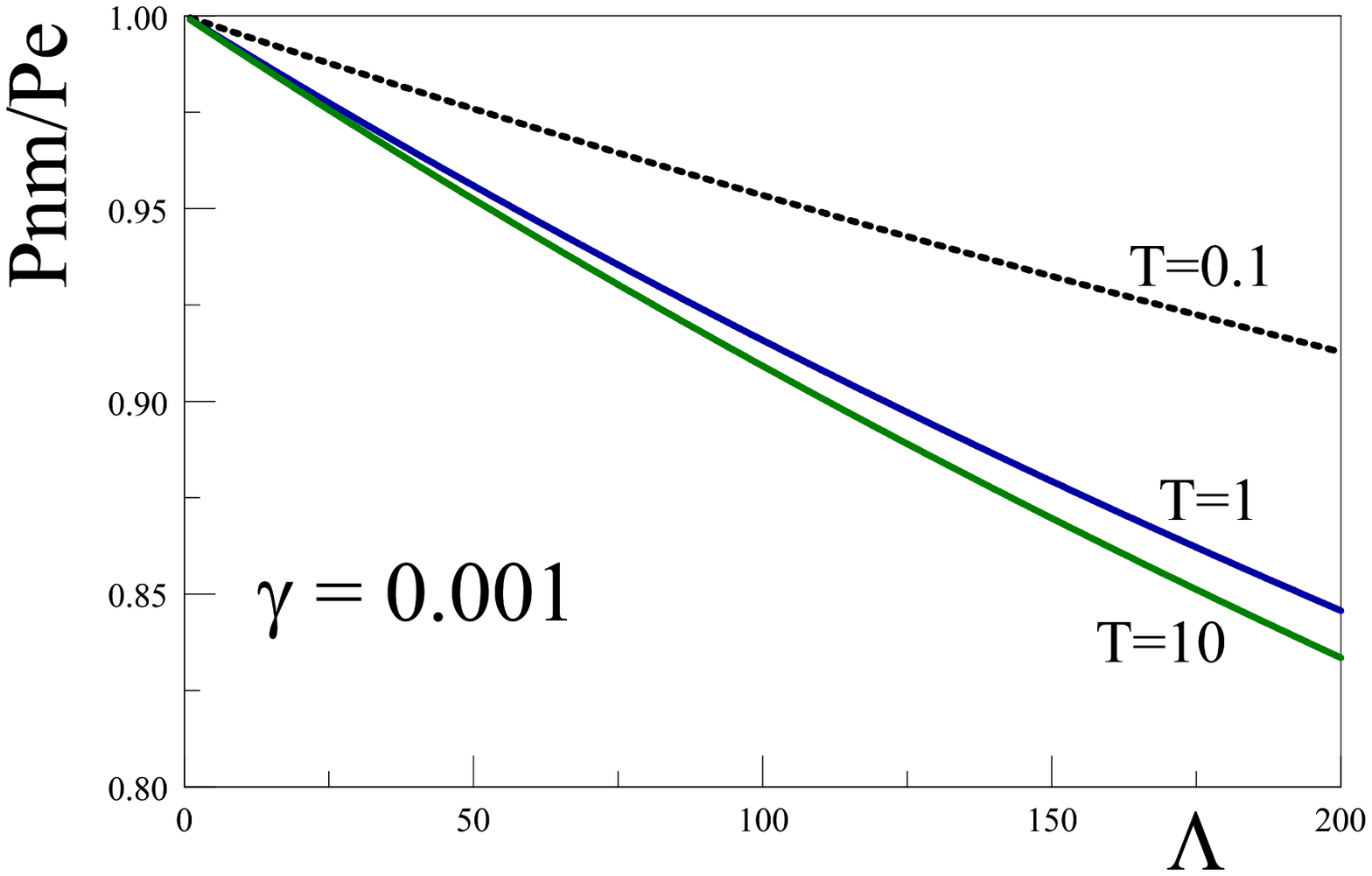}
\includegraphics[height=3.5in,width=3.5in,keepaspectratio=true]{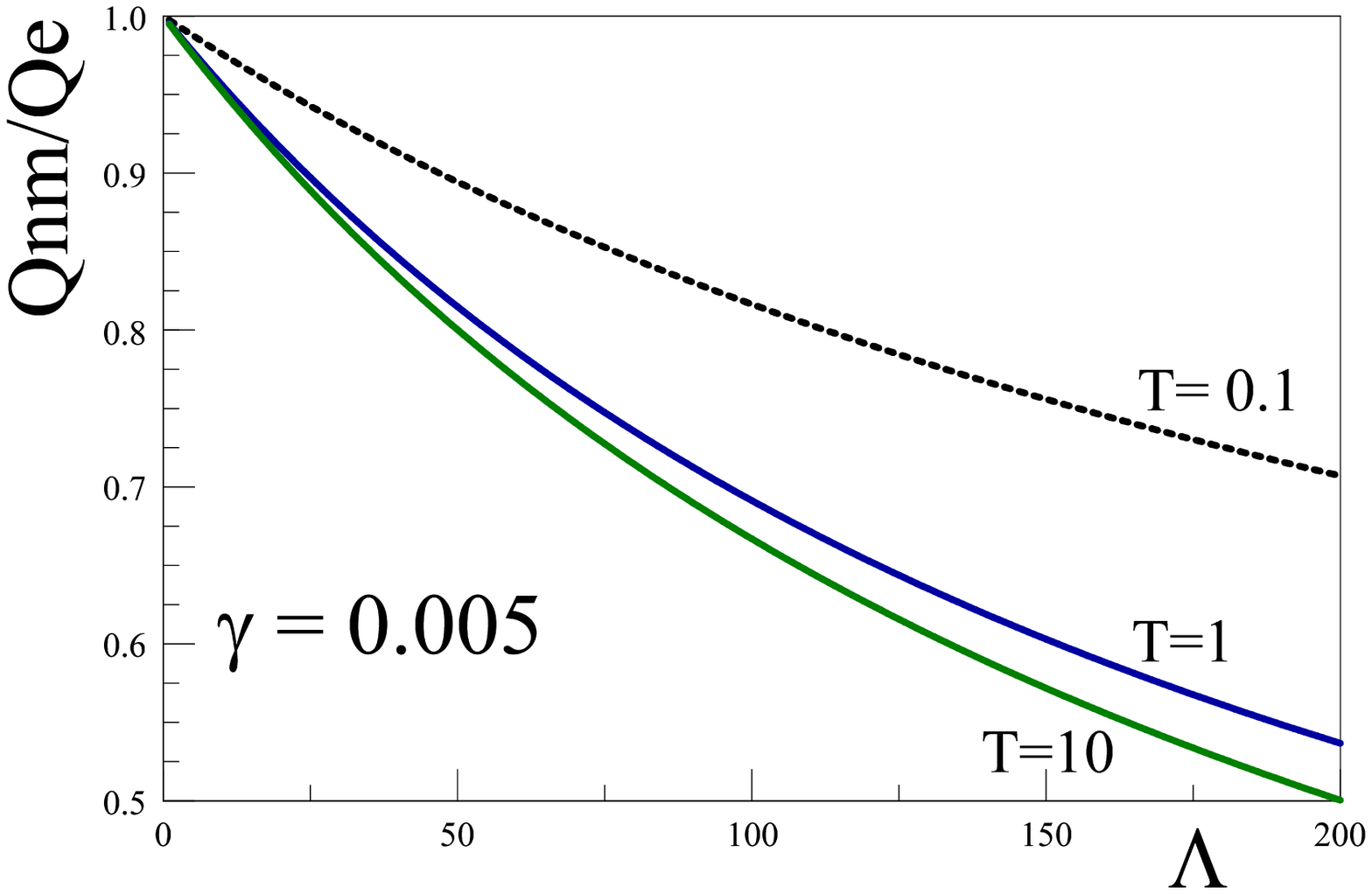}
\includegraphics[height=3.5in,width=3.5in,keepaspectratio=true]{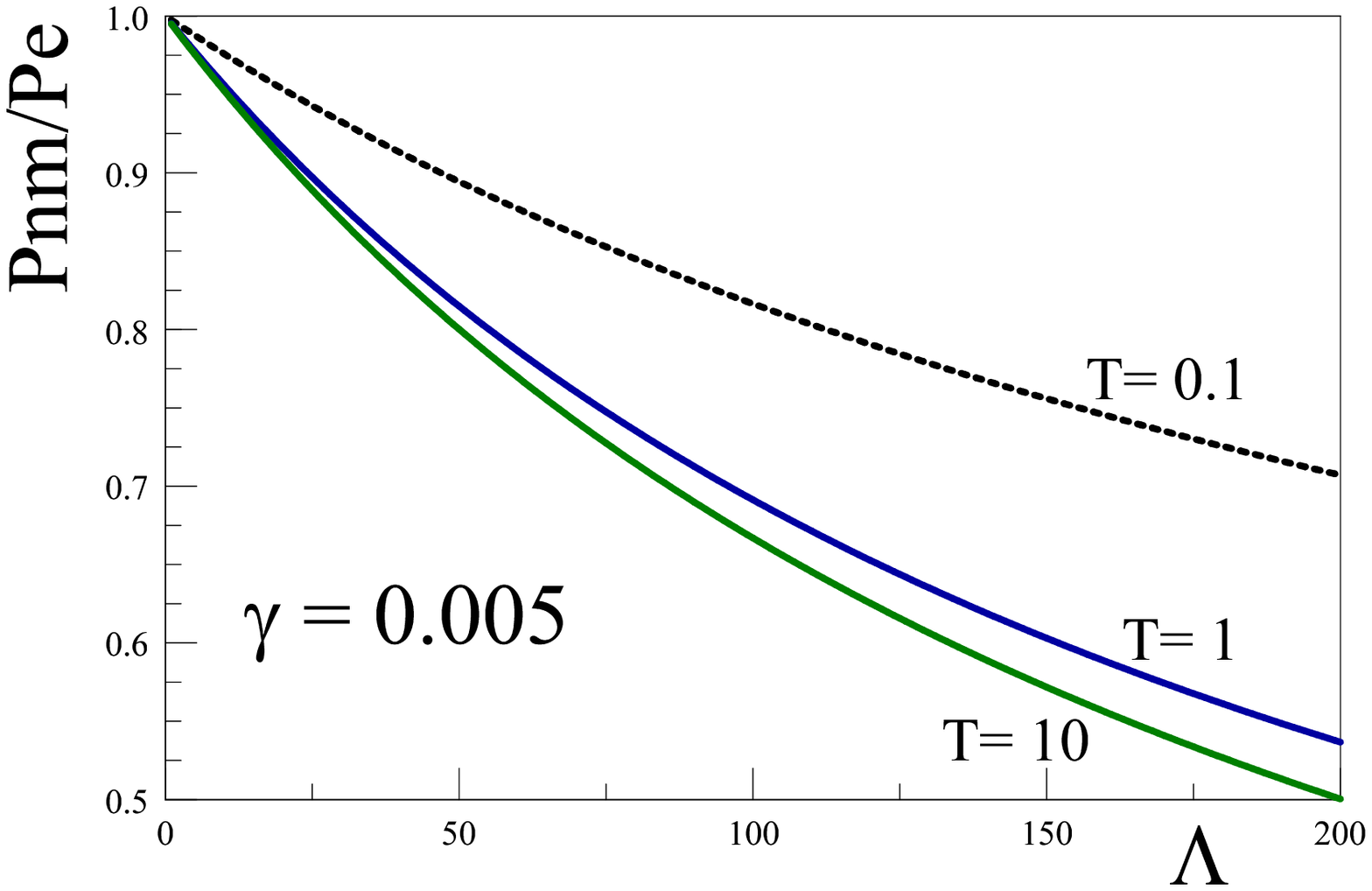}
\caption{ Non-Markovian case: the ratios $\frac{\langle q^2(\infty)\rangle_{NM}}{\langle q^2(\infty)\rangle_E} \equiv Qnm/Qe$, and $\frac{\langle p^2(\infty)\rangle_{NM}}{\langle p^2(\infty)\rangle_E} \equiv Pnm/Pe$, for $\gamma = 0.001~,~0.005$ vs. $\Lambda$ for various temperatures. $\gamma, \Omega, T,\Lambda$ are in units of $\Omega_R$.    }
\label{fig:nonmarkovratios}
\end{center}
\end{figure}

  \noindent  This analysis is confirmed by a numerical study of the Markovian and non-Markovian correlations displayed in figs. (\ref{fig:markovratios},\ref{fig:nonmarkovratios}) respectively. These figures display the ratios of the correlations to the exact results in the stationary limit   as a function of the bandwidth  for various values of temperature, both  in units of $\Omega_R$,  for $\gamma/ \Omega_R= 0.001; 0.005$ as representatives of the weak coupling regime. For example, in ref.\cite{groex} the Q-factor is $Q \equiv \Omega_R/\gamma \approx  215$. Weaker couplings yield smaller corrections both in the Markovian and non-Markovian cases. We confirm numerically that the largest contribution to the discrepancies with the exact results arise from the frequency renormalization (\ref{Oren}) whereas the logarithmic corrections are subleading for weak couplings in the range of bandwidths and temperatures that are physically relevant for the experimental settings with large $Q$ factors. The explicit comparison between the ratios in the Markovian and non-Markovian cases confirms the results (\ref{p2diff1}-\ref{q2diff2}) which, for weak couplings  are dominated by the first terms in the brackets, namely frequency renormalization.

         \vspace{2mm}

         \textbf{Counterterms?:} Within a   renormalized perturbative approach the term  $\Omega^2 \, q^2/2 $ in the system's Hamiltonian (\ref{Hsis}) is written as $\Omega^2_R \, q^2/2 + \delta \Omega^2 \, q^2/2$ where $\delta \Omega^2 = \Omega^2 -\Omega^2_R$ is a \emph{counterterm}  included in the interaction Hamiltonian and is required to cancel the $\Lambda$ dependence of the self-energy systematically in perturbation theory. In this approach, in the interaction picture  the system's degrees of freedom evolve in time  with $\Omega_R$.  Obviously this is not necessary in the \emph{exact} treatment with the Heisenberg equations of motion, where the full self-energy combines with the bare frequency to yield the exact solution solely in terms of the renormalized frequency. In the quantum master equation approach, the interaction Hamiltonian in the interaction picture is now
         \be H_I(t) = \frac{1}{2}\,\delta \Omega^2\,q^2(t) - q(t) \,\xi(t)\,, \label{HIcounter}\ee  instead of  (\ref{HSBint}), and $q(t)$ is given by  (\ref{qip}) with $\Omega \rightarrow \Omega_R$. The relation (\ref{Oren}) indicates that $\delta \Omega^2 = \gamma \Lambda$ is of \emph{second order} in the system-bath coupling.

         Following the steps leading   to Eq. (\ref{rhodotiter}) we immediately see that including the counterterm into the interaction Hamiltonian does \emph{not} resolve the discrepancy: taking the trace over the bath degrees of freedom, the first term in Eq. (\ref{rhodotiter}) no longer vanishes and yields a contribution that depends on the initial value, as it manifestly depends on $\rho_{SB}(0)$. Secondly, in perturbation theory the quadratic term in $H_I$ in (\ref{rhodotiter}) would yield a term proportional to $(\delta \Omega^2)^2 \propto \gamma^2$ whereas the system-bath interaction yields a term $\propto \gamma$. A term that is \emph{linear} in $\delta \Omega^2$ must vanish because it originates in a cross term linear in $\xi$ which vanishes upon tracing over the bath degrees of freedom. In conclusion, treating the difference between the renormalized and bare frequencies in terms of a counterterm Hamiltonian does not lead to a resolution of the discrepancies between the exact results from the Heisenberg-Langevin solution and those from the quantum master equation in the Born approximation with or without the additional Markovian assumption. Furthermore, such counterterm does not address the issue of the system-bath correlations neglected in the Born approximation. As discussed above, the Born-non-Markov quantum master equation   is exact within the factorization approximation and discrepancies with the exact result must originate in the factorization approximation. This is discussed in detail below.

\section{System-bath correlations.}\label{sec:SBcorrs}

The Born-non-Markov quantum master equation with memory only invokes the Born factorization, and,  as   shown above, within the factorization approximation \emph{only} the asymptotic long time limit yields the \emph{exact} equal time correlation functions under the assumption of a stationary state. Therefore, any discrepancy between the results of the Born-non-Markov quantum master equation and the exact solution of the Heisenberg-Langevin equations of motion for the asymptotic equal time correlation functions must originate in the factorization approximation. The approximation (\ref{factor}) neglects correlations between the system and the bath, which is justified when the system-bath interaction is weak, namely $\gamma/\Omega_R \ll 1$. Therefore discrepancies between the exact Heisenberg-Langevin and Born-non-Markov results are expected to be of $\mathcal{O}(\gamma/\Omega_R)$. However, the analysis above reveals that   the corrections are indeed of this order but enhanced by a large  factors $\Lambda/\Omega$ and  $\ln[\Lambda/T]$ for $\Lambda \gg T \gg \Omega_R,\gamma$.   These large enhancements suggest that the factorization implied by the Born approximation is missing important and large correlations between the system and the bath degrees of freedom. Although it is difficult to identify systematically the terms that are being neglected in the factorization (\ref{factor}), the exact solution of the  Heisenberg-Langevin equation (\ref{HLq})  allows us to quantify the reliability of the factorization by studying the \emph{correlation} between the system and bath variables.

The system's coordinate couples to the  \emph{collective} degree of freedom of the bath $B = \sum_k C_k Q_k$,   the correlation between the system's coordinate and this collective bath variable is precisely the interaction energy (\ref{SBcoup}). Therefore, we study the system-bath correlation by focusing on
\be  \langle ~H_{SB}(t) ~\rangle = -\langle \, q(t) B(t) \,\rangle ~~;~~ B(t) = \sum_k C_k Q_k(t) \,, \label{corre} \ee where $q(t)$ and $Q_k(t)$ are the \emph{exact} Heisenberg operators which are the solutions of eqns. (\ref{EOMq},\ref{EQMB}), namely (\ref{qoft}) and
\be B(t) = \xi(t) - \int^t_0 \Sigma(t-t') q(t') dt' \,.\label{Boft} \ee To obtain (\ref{Boft}) we used equations
(\ref{solB},\ref{noise}) and (\ref{sigma}) and the fact that the average in (\ref{corre}) is in the initial density matrix (\ref{rhototini}). Since $q(t)$ is given by (\ref{qoft}), we write
\be q(t) = q_\xi(t) + q_0(t) ~~;~~ B(t)= B_{\xi}(t) + B_0(t)\,, \label{Bsplit} \ee with
\be q_{\xi}(t) = \int^t_0 G(t-t') \xi(t') \,dt' \,,\label{qchi} \ee
\be q_0(t) = \dot{G}(t) q(0)+G(t) p(0) \,, \label{q0}\ee

\be B_{\xi}(t) = \xi(t)- \int^t_0 \Sigma(t-t') \int_0^{t'} G(t'-t'')\xi(t'') dt'dt'' \,, \label{bchi}\ee and
\be B_0(t) = -\int_0^t \Sigma(t-t')\Big[\dot{G}(t')q(0) + G(t') p(0) \Big]\,dt' \,. \label{Bo}\ee Because $\langle \langle \,\xi \rangle \rangle =0$, the cross  terms in the correlation function vanish, and we find
\be \langle \, q(t) B(t) \,\rangle = \langle \, q_0(t) B_0(t) \,\rangle + \langle \langle \, q_\xi(t) B_{\xi}(t) \, \rangle \rangle \,. \label{corr2}\ee Since $\langle \langle \xi \xi \rangle \rangle \propto \gamma$ and $\Sigma \propto \gamma$ clearly this correlation function --the interaction energy-- is proportional to $\gamma$.

 We are interested in the asymptotic limit $t \gg 1/\gamma$; in this limit the first term in Eq. (\ref{corr2}) vanishes since $G(t) \propto e^{-\gamma t/2}$, and only the second term survives. Therefore asymptotically,
\be \langle \, q(t) B(t) \,\rangle = \int^t_0 dt'\, G(t-t') \Big[ \langle \langle \,\xi(t') \xi(t) \,\rangle  \rangle   -    \int^t_0 \Sigma(t-t_1)\,\int^{t_1}_0 G(t_1-t_2) \langle \langle \,\xi(t') \xi(t_2) \,\rangle \rangle  dt_1 \, dt_2\Big] \,. \label{qgcorrasy} \ee The evaluation  of this expression is lengthy, but  because we want to understand the origin of the enhancement, we are only focused on extracting the leading terms in the limit $\Lambda \gg T \gg \Omega_R, \gamma$ . We find that the leading contributions for   $\Lambda \gg T, \Omega_R,\gamma$ in the asymptotic limit yield
\be \langle \, H_{SB}(\infty) \,\rangle = - {\langle \, q(t) B(t) \,\rangle}_{~~\overrightarrow{t\rightarrow \infty}}  ~~ - \gamma ~\Bigg\{
\frac{\Lambda}{2W} \, \Bigg[ \coth\Big[\frac{W+i\gamma/2}{2T}\Big]+\coth\Big[\frac{W-i\gamma/2}{2T} \Big]\Bigg] -\frac{1}{\pi} \ln\Big[ \frac{\Lambda}{2\pi T}\Big] \Bigg\} +\cdots \,, \label{qbasylam}\ee where the dots refer to terms that are subdominant for   $\Lambda \gg T, \Omega_R,\gamma$.  For $T \gg \Lambda \gg \Omega_R,\gamma$  the term $\ln(\Lambda/2\pi T)$ is replaced by $\ln(\Lambda/W)$. Again we note that the \emph{exact} expectation value of the asymptotic interaction energy does not depend directly on the bare frequency $\Omega$ but only on $\Omega_R$.

The  expectation value of the  interaction energy is proportional to $\gamma$ as expected but with proportionality factors that become very large in the limit $\Lambda \gg \Omega_R,\gamma$. The leading and next to leading order  terms  in the high temperature and weak coupling limits $T \gg \Omega_R \gg \gamma$ yield
\be \langle \, H_{SB}(\infty)\,\rangle = - 2T \, \Big(\frac{\gamma}{\Omega_R} \Big)\,\Big[\,\frac{  \Lambda}{\Omega_R} \, \Big]\,\Bigg\{ 1 -   \frac{\Omega^2_R}{2\pi \, T\,\Lambda}\,    \, \ln\Big[ \frac{\Lambda}{2\pi T}\Big] \,\cdots \Bigg\} \label{HiTHSB}\ee where we highlighted the dependence on the \emph{two} dimensionless ratios: $\gamma/\Omega_R$  that determines the  weak coupling  regime,  \emph{and} $\gamma \Lambda/\Omega^2_R$ along with the enhancement from the terms $\Lambda/\Omega_R$ and $\ln[\Lambda/T]$.

Although we cannot establish a direct link between the interaction energy and the corrections to the Born approximation (factorization)   in the time evolution of the density matrix, the $\Lambda$ dependence of the interaction energy suggests that these correlations are responsible for the discrepancy between the exact result for the equal time correlation functions from the Heisenberg-Langevin equations and those from the Born Markov and non-Markov quantum master equations.

\vspace{2mm}

\subsection{Energy flow}\label{sub:eflow}

The exact solution   also allows us to study how   energy is transferred between the bath and the system upon time evolution; the   system plus bath form a closed composite system, and energy is conserved. For the expectation value of the Hamiltonian in the initial state (\ref{rhototini}), consider the case $\rho_S(0) = (|0\rangle\langle 0|)_S$ so that
\be \langle \, H \, \rangle = \frac{\Omega}{2} + \sum_{k} \frac{W_k}{2}\, \coth\Big[\frac{W_k}{2T}\Big]  ~~;~~ \langle \, H_{SB} \, \rangle =0  \,.\label{totene}\ee This expectation value is time independent therefore the total change in the expectation value of $H_S(t)$ and $H_{SB}(t)$ compensate for the total change in the bath internal energy. Asymptotically, we find for changes in the system and interaction energies
\be \Delta E_S = \langle\, H_S(\infty) \,\rangle -\langle\, H_S(0) \,\rangle = \frac{1}{2} \langle p^2 \rangle + \frac{1}{2} (\Omega^2_R+\gamma \Lambda)\,\langle q^2 \rangle -\frac{1}{2}(\Omega^2_R+\gamma \Lambda)^{1/2} \,,\label{deltaEs}\ee
\be \Delta E_{SB} = \langle \, H_{SB}(\infty) \,\rangle \,,\label{deltaEI}\ee where $\langle p^2 \rangle, \langle q^2 \rangle, \langle\, H_S(\infty) \,\rangle$ are given by (\ref{p2fini},\ref{q2fini},\ref{qbasylam}) respectively, and we have written the system's energy in terms of $\Omega_R$ since the exact solution of the Heisenberg-Langevin equation depends solely on the renormalized frequency.

We note that the contribution from $\langle p^2 \rangle$ (Eq. (\ref{p2fini}) includes the contribution $\propto \, \gamma\, \ln[\Lambda/T]$ and the term $\gamma \Lambda  \, \langle q^2 \rangle$ includes a term $\propto \,\gamma \Lambda/W\, \coth[(\cdots)]$ from (\ref{q2fini}), these are precisely the terms shown in (\ref{qbasylam}).

 Energy conservation entails that the change in the internal energy of the bath is $\Delta E_B = -(\Delta E_S+\Delta E_{SB})$. The condition that the bath remains in thermal equilibrium at a fixed temperature implies that its internal energy $E_B \gg \Delta E_B$, although a specific calculation requires a definite dispersion relation for $W_k$. Consider the classical limit with equipartition and $E_B = \mathcal{N} T$, with  $\mathcal{N}$ being the number of degrees of freedom of the bath. The condition for $|\Delta E_B| / E_B \ll 1$ then requires that $\mathcal{N} \gg \gamma \Lambda /\Omega^2_R$. On physical grounds it is expected that $\Lambda \propto \mathcal{N}$ hence for sufficiently weak coupling  $\gamma/\Omega_R \ll 1$ this condition is likely to be always fulfilled and the bath remains in thermal equilibrium. A firmer assessment requires a model for the $W_k$ of the bath oscillators.

\subsection{Experimental settings}\label{sub:exp}

In experimental settings the strength of the coupling is measured by the mechanical quality factor $Q = \Omega/\gamma$, with a weak coupling regime corresponding to $Q \gg 1$. The stability condition (\ref{const}) implies that
\be \frac{\Lambda}{\Omega} < Q \,,\label{stable} \ee whereas the conditions for the validity of the Born approximation (\ref{bornconds}) imply \emph{both} $Q \gg 1$ and
\be \frac{\Lambda}{\Omega} \ll Q \,. \label{lQcondi} \ee On the other hand, after renormalization  an effective description of the dynamics of the system's degrees of freedom  should be insensitive on the dynamics of the high frequency components of the bath when $\Lambda \gg \Omega$. Therefore a ``coarse grained'' description of the dynamical evolution of the system's degrees of freedom   consistent with the Born approximation  requires that
\be 1 \ll \frac{\Lambda}{\Omega} \ll Q \,.  \label{winlam}\ee
Ref.\cite{groex} studies the amplitude response  of a mechanical resonator to thermal driving. The response is determined by the spectrum of the cavity output Eq. (4) in this reference,  which is identified with the frequency dependent integrand in Eq. (\ref{asyqtqtpnew}) in the high temperature limit with  $\coth[\omega/2T]\simeq 2T/\omega$.       The experimental settings in the study of micromechanical Brownian motion reported in Ref.\cite{groex} correspond to $\Omega = 2\pi \times  914 \,\mathrm{kHz}$ and $Q \simeq 215$ which implies a weak coupling to the environment. Fig. (2) in this reference shows the   response spectrum vs. frequency and prominently displays a resonance. Identifying the position of this resonance with $\Omega_R$ ($\Omega(\infty)$ in Ref.\cite{groex}) in the weak coupling limit (see Eq. (4) in \cite{groex}), inspection of the numerical value of the position of the resonance suggests that in this experiment
\be |\Omega^2-\Omega^2_R|/\Omega^2 \ll 1 \,. \label{rata}\ee \emph{If} the Drude-Ohmic model described this experiment, the result (\ref{rata}) would immediately yield $\gamma \Lambda /\Omega^2 \ll 1$. However, the main conclusion of this reference is that the spectrum of the bath $\sigma(\omega)$ fit within a broad window around the resonance yields $\sigma(\omega) \propto \omega^{-2.30 \pm 1.05}$. With the caveat that the measured spectrum is not of the Drude-Ohmic form, at least within this broad region of frequencies, it seems that the experimental values of the parameters in this experiment would justify using a Born (and Markov) approximation, although in Ref.\cite{groex} non-Markovianity is associated with a non-Ohmic (in this case sub-Ohmic) spectral density \emph{near the resonance}.

In Ref.\cite{teufel} a micromechanical oscillator is coupled to a microwave cavity to resolve the quantum vacuum fluctuations of the oscillator degrees of freedom. In this setting $\Omega = 2\pi\times 15.9 \,\mathrm{MHz}$ and the mechanical Q-factor is $Q = 10^5$. This is a very weakly coupled system, and although Ref.\cite{teufel} does not include an experimental value for $\Lambda$, or the difference between the bare and renormalized mechanical frequencies,  the constraint $\Lambda/\Omega \ll 10^5$ yields $\Lambda \ll 10^{7}\,\mathrm{MHz}$ leaving ample room to satisfy the conditions of validity for the Born approximation. Hence, the conditions for the validity of the the quantum master equation (either Markovian or non-Markovian) are likely fulfilled, and the Born quantum master equation may be well suited to study the dynamics in this case.

Recently Ref.\cite{norte} reported on novel designs for on-chip mechanical resonators for quantum optomechanical architectures with mechanical quality factors $Q \simeq 10^8$ which are sufficiently large to enter the quantum optomechanical regime at room temperature. This setting  corresponds to the micromechanical resonators being very weakly coupled to the environment,  with $\Omega \simeq 1\,\mathrm{MHz}$, again allowing a large range with $\Lambda/\Omega \ll Q$ for the validity of the Born approximation.

\section{  Non-Ohmic baths}\label{sec:nonohm}
Although we have focused on an Ohmic bath with a Drude spectral density, which allows an exact analytic solution of the Heisenberg-Langevin equations, the analysis presented above allows us to extrapolate some of the results to more general cases and draw more general conclusions. Following Ref.\cite{weiss} let us consider the general case
\be \sigma(\omega) = \gamma \omega \,\big| \frac{\omega}{\omega_0}\big|^{k-1}\,f(|\omega|/\Lambda)\,, \label{genspec}\ee where $\omega_0$ is a reference frequency taken to be different from the cutoff scale $\Lambda$ (in Ref.\cite{weiss} this latter scale is identified with a typical phonon frequency), and the dimensionless cutoff function $f(x)$ fulfills
\be f(x) \simeq 1 ~\mathrm{for}~ x \lesssim 1 ~~;~~ f(x) \rightarrow 0 ~\mathrm{for}~ x \gg 1 \,. \label{cutf}\ee The cases $k< 1,k=1, k> 1$ correspond to   sub-Ohmic, Ohmic and super-Ohmic respectively, and we assume that $\Lambda \gg \Omega_R, \gamma$.  The cases with  $k\leq -1$ yield an infrared divergent $\tilde{\Sigma}(s)$ and $\langle q^2 \rangle$ so we will focus on $k > -1$\footnote{The numerical fit to the data in ref.\cite{groex} was evidently performed within a range of frequencies near the position of the resonance; it does not include the region with $\omega \simeq 0$.}.

From the expression for the Green's function (in Laplace variable) given by (\ref{gofs}) the renormalization condition, which yields a (\emph{complex}) pole in weak coupling is \be \Omega^2_R = \Omega^2 + \widetilde{\Sigma} (s=i\Omega_R) \,,\ee where $\widetilde{\Sigma} (s) $ is given by the spectral representation (\ref{sigmaofs}).
Therefore we find that
\be (\Omega^2_R - \Omega^2)/\Omega^2 \simeq -\frac{\gamma \Lambda}{\Omega^2} \, \Big( \frac{\Lambda}{\omega_0}\Big)^{k-1} \,. \label{ratk}\ee The contribution to $\langle p^2 \rangle$ that diverges with $\Lambda$ is found to be
\be \langle p^2 \rangle \propto \frac{\gamma }{k-1}\, \Big( \frac{\Lambda}{\omega_0}\Big)^{k-1} \label{divp2gen}\ee (with a logarithmic divergence as $k \rightarrow 1$). Hence for \emph{super-ohmic} spectral densities the sensitivity to the cutoff scale is much stronger. This entails more stringent conditions for the validity of the Born-Markov  approximation beyond the usual weak coupling assumption $\gamma/\Omega \ll 1$; for example the condition (\ref{bornconds}) now becomes
\be \frac{\gamma \Lambda}{\Omega^2} \, \Big( \frac{\Lambda}{\omega_0}\Big)^{k-1} \ll 1 \,, \label{newcond}\ee which for $\Lambda \gg \omega_0$ requires a much weaker coupling and correspondingly larger $Q$.

\section{Conclusions}
The quantum master equation plays a fundamental role in   theoretical and experimental studies of quantum open systems. Its derivation often invokes   the Born and Markov approximations, the first corresponding to a factorization of the (reduced) density matrix of the system and that of the bath, which is  assumed to always remain at the initial value. The second neglects memory effects. Both approximations are usually justified for weak system-environment couplings. Motivated by the theoretical and experimental relevance of the quantum master equation, in this article we have studied the reliability of these approximations within the framework of quantum Brownian motion described by one mechanical oscillator (the system) linearly coupled to a   bath of harmonic oscillators in thermal equilibrium. This is an  important model for quantum open systems with definite experimental realizations in cavity optomechanics  and nano or micro-mechanical resonators. We solved exactly the Heisenberg-Langevin equations of motion for the degrees of freedom of the system and obtained the corresponding correlation functions. We considered a Drude-Ohmic spectral density for the thermal bath which allows us to analytically obtain the \emph{exact} correlation functions in the asymptotic stationary state as functions of temperature $T$, the  mechanical relaxation rate $\gamma$, bare $\Omega$ and renormalized $\Omega_R$ oscillator frequencies, and $\Lambda$ the bandwidth (cutoff) of the bath, assumed to satisfy $\Lambda \gg \Omega,\gamma$. The correlation functions depend on $\Omega_R$  but not directly on $\Omega$.  Within the Drude-Ohmic model,   stability and the existence of an
asymptotic steady state  requires the condition
\be \Omega^2_R >0 \Rightarrow \Omega^2 > \gamma \Lambda \,,\nonumber \ee and an effective long time description of the system's dynamics requires a separation between the time scale of the bath and that of the system, namely $\Lambda \gg \Omega_R$.

In the high temperature limit $T \gg \Omega_R,\gamma$ we recognize two different regimes: \textbf{i)} $T \gg \Lambda \gg \Omega_R, \gamma$, which is the classical regime where $\langle q^2 \rangle$ and $\langle p^2 \rangle$ obey classical equipartition (in terms of $\Omega_R$), and \textbf{ii)} $\Lambda \gg T \gg \Omega_R, \gamma$, in this regime there emerge corrections to $\langle p^2 \rangle \propto \gamma \ln\big[\Lambda/2\pi T\big]$, which, while perturbatively small for $\gamma/\Omega_R \ll 1$, are  enhanced by the logarithmic dependence on the bandwidth of the bath.

We then obtain the quantum master equation in the Born approximation and derive the equations of motion for expectation values and correlation functions both in the Markovian and non-Markovian case. In the latter case we introduce a systematic derivative expansion which yields the \emph{exact} correlation functions in the asymptotic stationary state within the Born approximation. Comparing the correlation functions obtained with the exact solutions of the Heisenberg-Langevin equation and those obtained in the Born approximation with and without the Markov approximation, we infer the conditions under which the Born approximation is reliable. We find that at least for the Drude-Ohmic case there are \emph{two} conditions
$$ \frac{\gamma}{\Omega} \ll 1 ~~;~~ \frac{\gamma \Lambda}{\Omega^2} \ll 1 \,. $$
The first is recognized as the usual weak coupling condition, but the second is a more stringent condition because for an effective description to emerge there must be wide separation between the time scale of the bath degrees of freedom $\propto 1/\Lambda$ and that of the system $\propto 1/\Omega$, namely $\Lambda \gg \Omega$. We have argued that differences in the asymptotic correlation functions between the exact results from the Heisenberg-Langevin solution, and those obtained with the Born-non-Markov (keeping the memory) quantum master equation must be a consequence of system-bath correlations being missed by the Born approximation. We studied the interaction energy as a \emph{proxy} for the system-bath correlations and found that the contributions that diverge as the cutoff $\Lambda \rightarrow \infty$ are \emph{precisely} the $\Lambda$ dependent terms in the difference between the correlation functions.

We have analyzed the above constraints as they apply to recent experimental settings and found that in these cases, the experimental values of parameters are consistent with both conditions being fulfilled, suggesting that the Born quantum master equation describes accurately the non-equilibrium dynamics in these experiments.

The study of  the Drude-Ohmic model allows us to draw more general conclusions for the cases of super-Ohmic, Ohmic and sub-Ohmic spectral densities. In particular, we find that the second constraint, involving the bandwidth of the bath becomes much more stringent in the super-Ohmic case, suggesting that non-Markovianity and system-bath correlations will play   very important roles in these cases, even for weak coupling.

Our results indicate that the conditions for the validity of the Born approximation with or without the Markov approximation, are    more stringent  than the usual condition on the quality factor of the mechanical oscillators $Q = \Omega/\gamma \gg 1$, which is equivalent to   weak coupling. The reliability of the Born quantum master equation as a tool to analyze   experimental results hinges crucially on the second condition $ \gamma \Lambda /\Omega^2 \ll 1$ or, alternatively, $\Lambda/\Omega \ll Q$. The fulfillment of this condition requires a detailed assessment of the experimental setting in each case.

Understanding the dynamics of decoherence is one of the main motivations in quantum open systems, the Heisenberg-Langevin approach being an exact method to study correlations should yield insight into the validity of Born-Markov quantum master equations usually implemented to extract decoherence rates. We expect to report on such study in a future article.

\acknowledgements The authors thank X.-L. Wu and A. Daley for fruitful  discussions. D. B. gratefully   acknowledges support from NSF through grant PHY-1506912. D.J. gratefully acknowledges the continued support of the Dietrich School of Arts and Sciences of the University of Pittsburgh.

\appendix
\section{Green's function} \label{app:GF}
Including the full expression for $\widetilde{\Sigma}(s)$ given by Eq. (\ref{laplasigma})
\be g(s) = \frac{s+\Lambda}{(s^2+ \Omega^2)(s+\Lambda)-\gamma \Lambda^2}\,. \label{fullgofs}\ee We write the denominator in (\ref{fullgofs}) as
\be (s^2+ \Omega^2)(s+\Lambda)-\gamma \Lambda^2 = (s-s_1)(s-s_2)(s-s_3) \label{prods}\ee where $s_{1,2,3}$ are the complex poles of $g(s)$, these obey the following conditions that can be read off the definition (\ref{prods}), namely
\bea s_1+s_2+s_3 & = & - \Lambda \label{sum1} \\
s_1s_2+s_3(s_1+s_2) & = & \Omega^2 \equiv \Omega^2_R + \gamma \Lambda \label{sum2} \\
s_1 s_2 s_3 & = & - \Lambda \, \Omega^2_R \label{sum3} \eea where $\Omega^2_R$ given by (\ref{Oren}) is the renormalized frequency.  In the formal limit $\Lambda \rightarrow \infty$  the function $g(s)$ is given by Eq. (\ref{gslarLam}) the denominator of which can be written as
\be s^2 + \Omega^2_R + \gamma s = (s-s_+) (s-s_-) ~~;~~ s_{\pm} = -\frac{\gamma}{2} \pm \sqrt{\Big(\frac{\gamma}{2}\Big)^2 - \Omega^2_R} \,.\label{denosmallgofs} \ee  We find the roots $s_{1,2,3}$ in the limit of $\Lambda \gg \Omega, \gamma$ by choosing
\be s_1 = s_+ + \mathcal{O}(1/\Lambda) ~~;~~ s_2 = s_- + \mathcal{O}(1/\Lambda) \label{approots}\ee and inserting into the constraints (\ref{sum1}-\ref{sum3}) we find
\be s_3 = -\Lambda +\gamma + \mathcal{O}(1/\Lambda) \,. \label{s3}\ee  Neglecting subleading terms in the limit $\Lambda \gg \Omega_R, \gamma$ we find the inverse Laplace transform
\be G(t) = \frac{\gamma}{\Lambda^2} \,e^{-\Lambda t} + e^{-\gamma t/2} \,\frac{\sin{Wt}}{W} + \mathcal{O}(\Omega_R/\Lambda^2, \gamma/\Lambda^2)~~;~~ W=  \sqrt{\Omega^2_R-\frac{\gamma^2}{4}}  \label{goftlead} \ee The first term in (\ref{goftlead}) not only is subleading ($\gamma/\Lambda \ll 1~~;~~ W/\Lambda \ll 1$) but can also be neglected after a very short time scale $t \simeq 1/\Lambda \ll 1/\gamma, 1/W$. Therefore the first term in (\ref{goftlead}) can be safely neglected and $G(t)$ coincides with the solution (\ref{goftlarLam}) valid for $\Lambda \gg \Omega_R,\gamma$.

\section{Integrals}\label{app:integrals}
With the Drude spectral density (\ref{ohm}) we find

\be i[g^>(\tau)-g^<(\tau)] = -\frac{i}{\pi}\int^\infty_{-\infty} \sigma(\omega)\,e^{i\omega \tau} = \gamma \Lambda^2 e^{-\Lambda \tau} ~~;~~ \tau > 0\,. \label{ggmingl}\ee
\be  \alpha(t) = \int^t_0 i[g^>(\tau)-g^<(\tau)]\,\cos(\Omega \tau) = \gamma \Lambda^2\,\mathrm{Re}\Bigg[ \frac{1-e^{-\Lambda \tau}\,e^{i\Omega \tau}}{\Lambda -i \Omega}\Bigg] \,, \label{alf}\ee
\be  \beta(t) = \int^t_0 i[g^>(\tau)-g^<(\tau)]\,\frac{\sin(\Omega \tau)}{\Omega} = \frac{\gamma \Lambda^2}{\Omega}\,\mathrm{Im}\Bigg[ \frac{1-e^{-\Lambda \tau}\,e^{i\Omega \tau}}{\Lambda -i \Omega}\Bigg] \,, \label{bet}\ee
\be g^>(\tau)+ g^<(\tau)= \frac{1}{\pi} \int^{\infty}_{-\infty} d\omega \, \sigma(\omega) \,\coth\Big[\frac{\omega}{2T}\Big] \,e^{i\omega \tau} \label{sumgs}\ee
with (\ref{ohm}) and
\be \coth\Big[\frac{\omega}{2T}\Big] = \frac{2T}{\omega} + 4T \sum_{l=1}^{\infty} \frac{\omega}{\omega^2 + \nu^2_l} ~~;~~ \nu_l = 2\pi l\,T ~~;~~ l=1,2,\cdots\label{integ}\ee the integral in (\ref{sumgs}) is carried out by residues closing the countour in the upper half plane. We find
\be g^>(\tau)+ g^<(\tau) = i\gamma \Lambda^2 \,\coth\Big[\frac{i\Lambda}{2T}\Big]\,e^{-\Lambda \tau}-4 \gamma \Lambda^2 \,T \sum_{l=1}^{\infty} \frac{\nu_l}{\Lambda^2 - \nu^2_l} \,e^{-\nu_l \tau} \label{sumgsfin}\ee The integral
\be \int^t_0 d\tau (g^>(\tau)+ g^<(\tau))\,e^{i\Omega \tau} = \gamma \Lambda^2 \Bigg\{ i\coth\Big[\frac{i\Lambda}{2T}\Big] ~\frac{[1-e^{-\Lambda t}\,e^{i\Omega t}]}{\Lambda^2+\Omega^2} \big(\Lambda + i\Omega\big) - 4 T \sum_{l=1}^{\infty} \frac{\nu_l}{\Lambda^2 - \nu^2_l}~\frac{[1-e^{-\nu_l t}\,e^{i\Omega t}]}{\nu_l^2+\Omega^2} \, (\nu_l + i\Omega) \Bigg\}\,. \label{intet}\ee In the asymptotic long time limit we find
\be f(\infty) = \int^\infty_0 (g^>(\tau)+ g^<(\tau))\,\frac{\sin(\Omega \tau)}{\Omega} = 2\gamma \Bigg\{ -\frac{T}{\Lambda} - \frac{1}{\pi} \mathrm{Re} \Big[\Psi\Big(\frac{\Lambda}{2\pi T} \Big)-\Psi\Big(\frac{i\Omega}{2\pi T} \Big)\Big]\Bigg\} \label{finfty} \ee where $\Psi$ is the di-Gamma function, and for $\Lambda \gg \Omega$ we find
\be h(\infty) =  \int^\infty_0 (g^>(\tau)+ g^<(\tau))\, {\cos(\Omega \tau)} = \gamma \Omega\,\coth\big[\frac{\Omega}{2T}\big] \,.\label{hinfty} \ee

    Using the result (\ref{sumgsfin}) we find for the symmetrized correlation function
    \be \frac{1}{2} \,\langle \langle \Big(\xi(t_1) \xi(t_2)+ \xi(t_2) \xi(t_1)\Big)  \rangle \rangle =  \gamma \Lambda^2 \Bigg[ \frac{i}{2}\,\coth\Big[\frac{i\Lambda}{2T}\Big]\,e^{-\Lambda |t_1-t_2|}-  \frac{1}{\pi} \,         \sum_{l=1}^{\infty} ~\frac{l~~e^{-2\pi l\,T\, |t_1-t_2|}}{\Big(\frac{\Lambda}{2\pi T}\Big)^2 - l^2} ~ \Bigg]   \,. \label{symecor}\ee In the classical limit $T \gg \Lambda$ this correlation function becomes
    \be  \gamma T \Lambda \,e^{-\Lambda |t_1-t_2|}-\frac{1}{\pi}\gamma \Lambda^2 \,\ln\Big[1- e^{-2\pi\,T\, |t_1-t_2|}\Big] \,.\label{classsymenoise}\ee For $|t_1-t_2| \gg 1/2\pi T$ the second term can be neglected and in the formal limit $\Lambda \rightarrow \infty$ the first term
    $$ \rightarrow \gamma T \delta(t_1-t_2)\,, $$ which is the usual form for the classical correlation function white noise.

    The functions $F(\Lambda/T,\Omega_R/T,\gamma/T)$ in eqn. (\ref{q2fini}) and $I(\Lambda/T,\Omega_R/T,\gamma/T)$  in eqn. (\ref{p2fini}) are  given by
    \be F(\Lambda/T,\Omega_R/T,\gamma/T) = \frac{1}{4\pi^3}~\Big(\frac{\Lambda}{2\pi T}\Big)~\sum_{l=1}^{\infty}\Bigg[\frac{l}{l^2-\Big(\frac{\Lambda}{2\pi T}\Big)^2} \,\frac{1}{\Big(l^2+\Big(\frac{\Omega_R}{2\pi T}\Big)^2 \Big)^2-l^2\,\Big(\frac{\gamma}{2\pi T}\Big)^2} \Bigg]\,.  \label{FunFu}\ee

\bea I(\Lambda/T,\Omega_R/T,\gamma/T) & = &   -\frac{1}{\pi}\, \Big(\frac{\Lambda}{2\pi T}\Big)^2~\sum_{l=1}^{\infty}\Bigg[\frac{l}{l^2-\Big(\frac{\Lambda}{2\pi T}\Big)^2} \,\frac{l^3}{\Big(l^2+\Big(\frac{\Omega_R}{2\pi T}\Big)^2 \Big)^2-l^2\,\Big(\frac{\gamma}{2\pi T}\Big)^2} \Bigg] \nonumber \\ &-&   \frac{i}{2}\,\coth\big[\frac{i\Lambda}{2T} \big] \,.  \label{FunIu}\eea

\end{document}